\RequirePackage{lineno}
\documentclass[twocolumn,showpacs,superscriptaddress,amsmath,amssymb,nofootinbib]{revtex4-2}
\usepackage{graphicx}
\usepackage{subfigure}
\usepackage{subcaption}
\usepackage{epsfig}
\usepackage{overpic}
\usepackage{dcolumn}
\usepackage{ulem}
\usepackage{bm}
\usepackage{color}
\usepackage{lineno}
\usepackage{xspace}
\usepackage{multirow}
\usepackage{epstopdf}
\usepackage{xcolor}
\usepackage{soul}
\usepackage{verbatim}
\usepackage{enumitem}
\usepackage{todonotes}
\usepackage{cancel}
\usepackage[toc,page]{appendix}
\usepackage[colorlinks,allcolors=black]{hyperref}

\usepackage{diagbox}

\newcommand{\ihep}{\affiliation{State Key Laboratory of Nuclear Physics and Technology, Institute of Quantum Matter, South China Normal University, Guangzhou 510006, China
}}

\newcommand{\moe}{\affiliation{Key Laboratory of Atomic and Subatomic Structure and Quantum Control (MOE), Guangdong-Hong Kong Joint Laboratory of Quantum Matter, Guangzhou 510006, China}}

\newcommand{\iqm}{\affiliation{Guangdong Basic Research Center of Excellence for 
Structure and Fundamental Interactions of Matter, Guangdong 
Provincial Key Laboratory of Nuclear Science, Guangzhou 
510006, China}}

\newcommand{\scnt}{\affiliation{Southern Center for Nuclear-Science Theory (SCNT), Institute of Modern Physics, Chinese Academy of Sciences, Huizhou 516000, Guangdong Province, China}}

\newcommand{\SJTU}{\affiliation{State Key Laboratory of Dark Matter Physics, Shanghai Key Laboratory for Particle Physics and Cosmology, Key Laboratory for Particle Astrophysics and Cosmology (MOE), Physics-AI institute, School of Physics and Astronomy, Shanghai Jiao Tong University, Shanghai 200240, China}}

\begin{document}
\include{def-com}
\title{Resolving the CP Asymmetry Puzzle in $B$ Decays with Unitarized Final-State Interactions}
\ihep
\moe
\iqm
\SJTU
\scnt



\author {Xin-Yue Hu}
\ihep
\moe
\iqm

\author {Pengyu Niu}
\email{niupy@m.scnu.edu.cn, corresponding author}
\ihep
\iqm

\author {Qian Wang}
\email{qianwang@m.scnu.edu.cn, corresponding author}
\ihep
\iqm
\scnt
\author{Wei Wang}
\email{wei.wang@sjtu.edu.cn, corresponding author}
\SJTU
\scnt

\date{\today}

\begin{abstract}
A reliable description of direct CP asymmetries in hadronic $B$ decays, among the most sensitive probes of the CKM mechanism and of physics beyond the Standard Model, remains unresolved. Conventional factorization approaches adopt distinct treatments for the strong phases originating from long-distance QCD interactions, leading to predictions that differ by factors of several for identical decay channels. Existing final-state interaction (FSI) models are limited to one-loop approximations that break unitarity and rely on channel-specific phenomenological cutoffs, severely restricting the predictive capability. 
We introduce a unitarized FSI framework based on the Lippmann-Schwinger equation solved to all orders, restoring unitarity that is manifestly broken in one-loop treatments. Using the coupled $D\bar{D}$ system, we show that the interaction kernel can be derived from chiral effective field theory constrained by heavy-quark spin symmetry, and low-energy constants are fixed by the $\gamma\gamma\to D\bar{D}$ cross-section data from BaBar, leaving no adjustable parameter in the FSI sector when applied to $B$ decays. The resulting predictions for CP asymmetries and branching fractions show excellent agreement with the available  experimental data. 
A defining consequence is that CP asymmetries in the pure-annihilation channels $\bar{B}^0\to D^0\bar{D}^0$ and $\bar{B}^0\to D_s^+D_s^-$, identically zero in any short-distance treatment, are dramatically enhanced by FSI, and a  measured nonzero asymmetry in these modes is therefore a direct experimental signature of long-distance dynamics. We present definite predictions for the partial widths and CP asymmetries of all yet-unmeasured channels, establishing final-state interactions as a predictive, data-driven ingredient of the Standard Model with direct implications for CKM parameter extraction and BSM searches. 
\end{abstract}
\maketitle


{\it Introduction.} Non-leptonic $B$ decays provide a critical probe of CP violation and the CKM mechanism of the Standard Model (SM)~\cite{Sakharov:1967dj,UTfit:2022hsi}, and their direct CP asymmetries are among the most sensitive observables for physics beyond the SM.
The theoretical description of these decays rests on factorization theorems~\cite{Beneke:1999br,Beneke:2000ry,Keum:2000wi,Lu:2000em,Bauer:2000yr,Bauer:2004tj}, which become exact in the heavy-quark limit $m_b\to\infty$. Yet the strong phases that drive CP violation are dominated by power-suppressed long-distance effects that factorization excludes by construction. The predictions of the factorization methods for direct CP asymmetries can significantly  differ in benchmark channels~\cite{Beneke:2003zv,Cheng:2009mu,Cheng:2009cn,Chai:2022ptk,Zou:2015iwa,Williamson:2006hb,Wang:2008rk}. This situation precludes a reliable extraction of CKM parameters and weakens the sensitivity to new physics.

Final-state interactions (FSI) are a long-recognized source of these missing strong phases~\cite{Cheng:2004ru,Bediaga:2020qxg}.
In charm decays, coupled-channel FSI effects have been shown to dramatically alter CP asymmetries~\cite{Pich:2023kim,Bediaga:2022sxw}, and in $B$ decays to light mesons, rescattering has been invoked to resolve the long-standing $K\pi$ puzzle~\cite{Falk:1998wc,Donoghue:1996hz}.
Nevertheless, existing FSI studies of $B$ decays suffer from two fundamental limitations.
First, they rely on one-loop approximations~\cite{Cheng:2004ru,Mohammadi:2011zz}, which assume that final-state particles rescatter only once---a restriction with no physical justification and which manifestly breaks unitarity.
Second, the phenomenological cutoffs that regularize the loop integrals vary from channel to channel and cannot be predicted {\it a priori}~\cite{Cheng:2004ru}, rendering FSI an untestable theoretical ingredient rather than a predictive tool, and thus a reliable understanding of CP asymmetries in $B$ decays is not established. 

In this Letter, we construct a systematically improvable, data-driven FSI framework for $B_{(s)}$ decays. We take the $B\to D\bar{D}$ channels as the example, where the final-state interaction is dominated by $S$-wave $D^0\bar{D}^0$--$D^+D^-$--$D_s^+D_s^-$ rescattering.
The $D\bar{D}$ $T$-matrix is obtained non-perturbatively by solving the Lippmann-Schwinger equation $T = V + V G T$ with a contact potential $V$ derived from chiral effective field theory and constrained by heavy-quark spin symmetry.
Every low-energy constant in $V$ is fixed independently by the $\gamma\gamma\to D\bar{D}$ cross-section data from BaBar~\cite{BaBar:2010jfn}, leaving {\it no} adjustable parameter for the FSI sector when applied to $B$ decays.
Short-distance decay amplitudes are parameterized within the operator product expansion by only three complex constants, determined in a global fit to the measured branching fractions.
With every FSI degree of freedom fixed by independent scattering data, the CP asymmetries of yet-unmeasured channels become genuine predictions.
Crucially, channels such as $\bar{B}^0\to D^0\bar{D}^0$ and $\bar{B}^0\to D_s^+D_s^-$ receive contributions solely from annihilation topologies, for which no asymmetry can arise without rescattering.
The use of HQSS-constrained chiral EFT reduces the entire FSI sector to just two low-energy constants, fixed entirely from $\gamma\gamma\to D\bar{D}$ data, so that the rescattering dynamics remains data-driven rather than parametric when applied to $B$ decays.
A nonzero CP asymmetry measurement in these modes is therefore an unambiguous signature of long-distance strong phases, which cannot be accommodated  with short-distance mechanism.

\begin{figure}[htp!]
    \centering
    \includegraphics[width=0.96\linewidth]{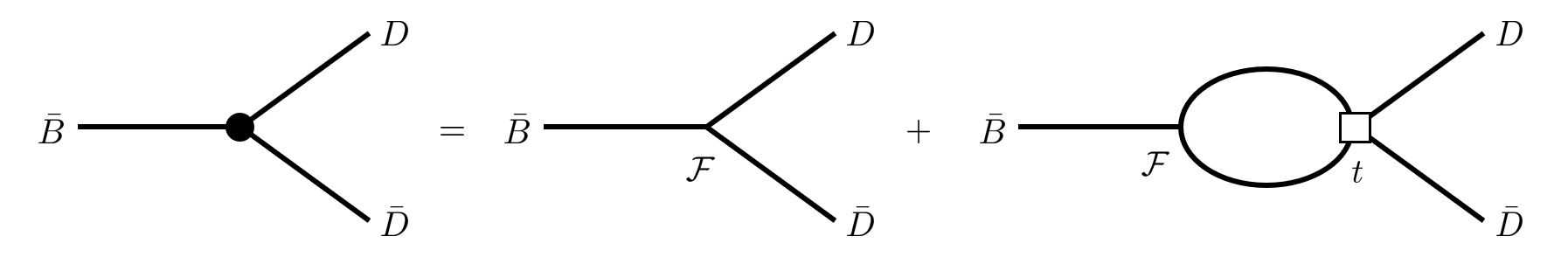}
    \caption{Schematic of $\bar{B}$ decay to a charmed-meson pair including FSI. $\mathcal{F}$ labels the bare vertices and t the rescattering amplitude.}
    \label{fig:framework}
\end{figure}

{\it Framework of unitarized FSI.}~~
As shown in Fig.~\ref{fig:framework}, 
the bottom meson $B=(B^-,\bar{B}^0,\bar{B}_s^0)$ decay to a pair of charmed mesons $F=D_i\bar{D}_j$, with $i,j=1,2,3$ light flavor index for $u,d,s$ quark. Subsequently, the $D\bar{D}$ pairs 
undergo final-state interactions (FSI), 
which are described by a unitarized rescattering amplitude $t$.   

Using the Operator Product Expansion (OPE),
decay amplitude is factorized into short-distance and long-distance parts. The short-distance effects, above the energy scale 
$\mu$, are encoded in Wilson coefficients
$C_i(\mu)$. The long-distance effects, below
the energy scale $\mu$, are contained in the hadronic matrix elements
$\left\langle F \right|\mathcal{O}_{i}(\mu)\left| B\right\rangle$.
Here $\mu\sim m_b$ is the renormalization scale~\cite{Buras:1998ra}. 
In this process, there are ten
operators $\mathcal{O}_{1-10}$~\cite{Buras:1998ra,Ciuchini:1997hb}, among which, only the current-current operators $\mathcal{O}_{1,2}$ are retained, because the Wilson coefficients $C_{3-10}$ are suppressed by $\frac{\alpha_s}{12\pi}\ln (\frac{m_t^2}{\mu^2})$, due to the short-distance contribution of the virtual top quark~\cite{Ciuchini:1997hb}.  In this case,
the decay amplitude can be written as 
\begin{equation}
    \mathcal{M}(B\to F)=\frac{G_F}{\sqrt{2}}V_{qb}^{*}V_{qq^{\prime}}\sum_{n=1,2}C_{n}(\mu)\left\langle B\right|\mathcal{O}_{n}^{qq^{\prime}}(\mu)\left| F\right\rangle,
    \label{eq:prod_amplitude}
\end{equation}
where $G_{F}$ is the Fermi constant, $V_{ij}$ are Cabibbo-Kobayashi-Maskawa (CKM) matrix elements, and 
\begin{equation}
\begin{aligned}
&\mathcal{O}_{1}^{qq^{\prime}}= (\bar{q^{\prime}}_{\alpha}q_{\beta})_{V-A}(\bar{q}_{\beta}b_{\alpha})_{V-A},\\
    &\mathcal{O}_{2}^{qq^{\prime}}= (\bar{q^{\prime}}_{\alpha}q_{\alpha})_{V-A}(\bar{q}_{\beta}b_{\beta})_{V-A},
\end{aligned}
    \label{eq:operators}
\end{equation}
with $q=(u,c)$, $q^{\prime}=(d,s)$, and the vector-axial vector current $(q_{1}q_2)_{V-A}=q_1\gamma_{\mu}(1-\gamma_5)q_2$. $\alpha$ and  $\beta$ are color indexes. 
From topological considerations, the amplitude can be parameterized by seven complex parameters (the details can be  found in the Supplemental Material), which 
reduce to three independent ones, i.e. $T$, $A$, and $P$, after neglecting CKM-suppressed contributions.

When FSI is included,
the decay width becomes
\begin{equation}
    \begin{aligned}
        \Gamma(\bar{B}^i\to D^{j}\bar{D}^{j}&)=\frac{\sqrt{M_i^2-4m_{j}^2}}{16\pi M_{i}^2}\\
        &\times|\mathcal{F}_{ij}+\sum_{k}\mathcal{F}_{ik}G^{\Lambda}_{k}(M_i^2)t_{kj}(M_{i}^2)|^2,
    \label{eq:decay width}
    \end{aligned}
\end{equation}
where $M_i$ denotes the mass of the initial $\bar{B}^i$, and $m_{j}$ denotes the mass of $D^j$.
$\mathcal{F}_{ij}$ represents the direct production amplitude of the   $D^j\bar{D}^j$ from the $\bar{B}^i$ (see Tab.~\ref{tab:decay channel}).
$G_k^\Lambda$ is the two-body loop function and $t_{kj}$ is the unitarized rescattering amplitude, which will be discussed in the next section. 

{\it Unitarized $D\bar{D}$ scattering.}~~
The predictive description of direct CP asymmetries in hadronic $B$ decays is limited by long-distance strong phases,
which are absent in factorization at leading power and are commonly modeled through one-loop rescattering \cite{Cheng:2004ru,Geng:2025yna,Duan:2024zjv,Bediaga:2018wml}. Such treatments are intrinsically incomplete: they allow only a single rescattering and violate unitarity at the amplitude level. Since direct CP asymmetries are controlled by the interference of weak and strong phases, a unitary treatment of final-state interactions (FSI) is essential.

Recently, the BaBar Collaboration measured $\gamma\gamma\to D\bar{D}$ cross sections~\cite{BaBar:2010jfn}.  We extract the low-energy constants from the $\gamma\gamma\to D\bar{D}$ cross sections. Here, we only consider the $S$-wave $D\bar{D}$ interactions in the $\gamma\gamma\to D\bar{D}$ process, with the amplitude of the $i$-th channel reads as
\begin{equation}
    \mathcal{M}(\gamma\gamma\to D^i\bar{D}^i)=\mathcal{C}_jG^{\Lambda}_j(s)t_{ji}(s)+\mathcal{C}_i,
\end{equation}
where $\mathcal{C}_i$ is the bare production amplitude for the $i$-th channel, with $i=1,2,3$ for the $D^0\bar{D}^0$, $D^+D^-$, $D_s^+D_s^-$ channels, respectively. Here the bare production amplitudes are parameterized as $\mathcal{C}_1=r\mathcal{U}$, $\mathcal{C}_2=\mathcal{U}$, $\mathcal{C}_3=\mathcal{U}$, with $\mathcal{U}$ the bare production amplitude for  $\gamma\gamma$ directly create $D^+D^-$ or $D_s^+D_s^-$, and $r$ the relative strength between the neutral channel and the charged one. In this case,
the cross section of the $i$-th channel is expressed as
\begin{equation}
    \sigma_i(s)=\frac{\sqrt{s(s-4m_{i}^2)}}{16\pi s^2}|\mathcal{M}(\gamma\gamma\to D^i\bar{D}^i)|^2.
\end{equation}
The experimental resolution is also considered in our calculation. The details can be found in the Supplemental Material.

{\it CP asymmetries in unitarized FSI.}~~The direct CP asymmetry is defined as
\begin{equation}
    \mathcal{A}_{\mathrm{CP}}(\bar{B}\to F)\equiv \frac{\Gamma(\bar{B}\to F)-\Gamma(B\to F)}{\Gamma(\bar{B}\to F)+\Gamma(B\to F)}.
    \label{eq:ACP}
\end{equation}
Without FSI, the channels $\bar{B}^0\to D^0\bar D^0$ and $\bar{B}^0\to D_s^+D_s^-$ are pure-annihilation modes [Table~\ref{tab:decay channel}] and therefore satisfy
\begin{eqnarray}
    \mathcal{A}_{\mathrm{CP}}(\bar{B}^0\to D^0\bar{D}^0)
    =
    \mathcal{A}_{\mathrm{CP}}(\bar{B}^0\to D_s^+D_s^-)=0,
    \label{eq:ACPwoFSI}
\end{eqnarray}
in any short-distance treatment. This exact null test is one of the central results of the short-distance framework. By contrast, once coupled-channel rescattering is included, nontrivial strong phases are generated and the asymmetries become
\begin{equation}
    \begin{aligned}
    \mathcal{A}_{\mathrm{CP}}(\bar{B}^0\to D^0\bar{D}^0)&\simeq f(\delta_p+\Delta_{12}^1+\Delta_{32}^1-\Delta_{22}^1),\\
    \mathcal{A}_{\mathrm{CP}}(\bar{B}^0\to D^+D^-)&\simeq f(\delta_p+\Delta_{12}^{\prime}+\Delta_{32}^{\prime}),\\
    \mathcal{A}_{\mathrm{CP}}(\bar{B}^0\to D_s^+D_s^-)&\simeq f(\delta_p+\Delta_{12}^3+\Delta_{32}^3-\Delta_{22}^3),
    \end{aligned}
\label{eq:ACP1}
\end{equation}
where $f\equiv {2R\sin\gamma}/(1+R^2-2R\cos\gamma)$ with $R\equiv |Y_{ud}/Y_{cd}|$ and $\gamma\equiv \arg(-Y_{ud}/Y_{cd})$ and the $\Delta_{ij}^k$ encode the rescattering corrections. Their explicit expressions are given in the Supplemental Material. The channels $B^-\to D^0D^-$ and $\bar{B}_s^0\to D_s^+D^-$, for which FSI do not induce new interference structures, satisfy
\begin{equation}
    \mathcal{A}_{\mathrm{CP}}(B^-\to D^0D^-)=f\delta_p,
    \label{eq:ACPinfit}
\end{equation}
and are therefore especially useful for constraining the short-distance phase $\delta_p$ directly from data.

To quantify the role of rescattering, we perform the three fits summarized: Scheme I includes only the short-distance amplitudes $(T,P,A)$ while fitting branching fractions, Scheme II adds the unitarized FSI additionally, and Scheme III additionally includes the measured direct CP asymmetries in the channels insensitive to FSI. This setup allows us to separate cleanly the impact of rescattering on rates from its impact on CP-violating observables.

 The short-distance-only description, Scheme I, gives a poor fit with $\chi^2/{\rm d.o.f}=2.30$, indicating that a purely quark-level treatment is insufficient. Once the unitarized FSI is included, the fit quality improves dramatically to $\chi^2/{\rm d.o.f}=1.15$ in Scheme II. Incorporating the measured CP asymmetries in the FSI-insensitive channels further constrains $\delta_p$ and leads to $\chi^2/{\rm d.o.f}=1.00$ in Scheme III. This pattern is highly nontrivial: FSI are not introduced as a flexible phenomenological correction, but as an independently calibrated dynamical ingredient whose inclusion is decisively favored by the data.

A second notable feature is the stability of the fitted short-distance amplitudes across the three schemes (see the Supplementary Materials). The magnitudes of $T$, $A$, and $P$ remain nearly unchanged, which shows that the improvement is driven primarily by the long-distance rescattering dynamics rather than by a reshuffling of the short-distance sector. Moreover, the fitted values are compatible in order of magnitude with independent estimates from pQCD \cite{Zhou:2015jba,Zhou:2024qmm,Ou-Yang:2025ije} and pole models\cite{Fu-Sheng:2011fji,Colangelo:2003sa,Colangelo:2002mj}, supporting the separation between short- and long-distance contributions.
\begin{figure}[htp!]
    \centering
    \includegraphics[width=0.96\linewidth]{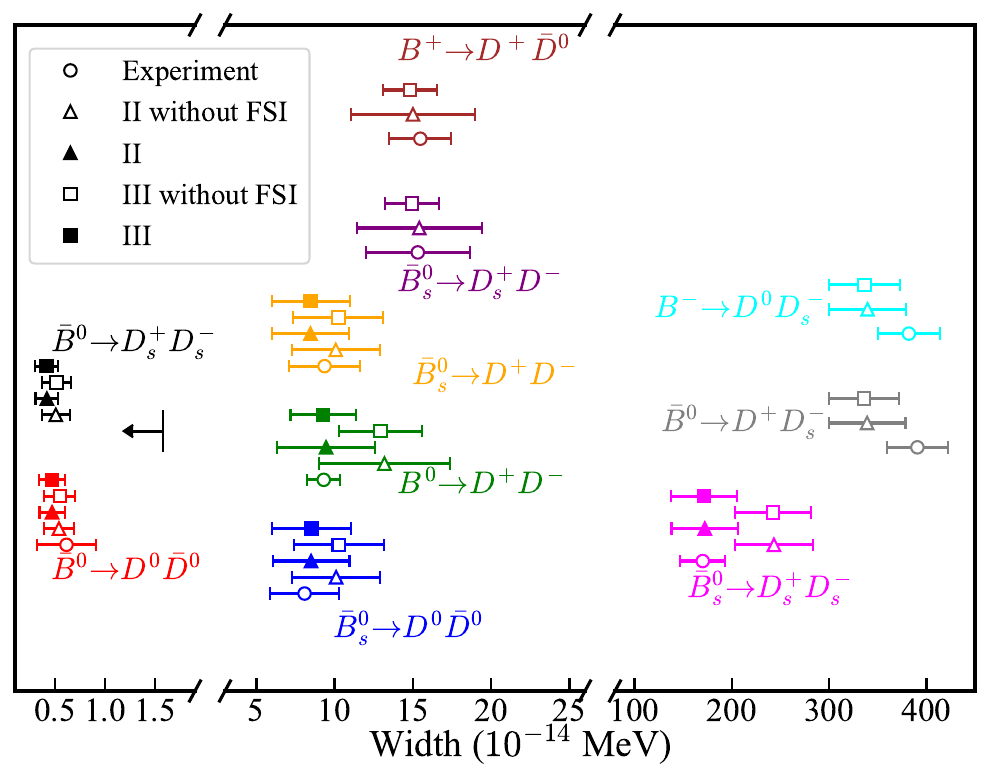}
    \caption{Comparison of the predicted partial widths with experiment~\cite{Belle:2012mef,Zupanc:2007pu,LHCb:2013sad,Belle-II:2023bps,LHCb:2017knt,LHCb:2014qsd,Belle-II:2024xtf,Belle:2008doh}. Hollow circles denote data. Solid triangles and squares correspond to Schemes II and III including FSI, while hollow markers show the corresponding results with FSI switched off but without refitting. In the $\bar{B}^0\to D_s^+D_s^-$ channel, the vertical bar with a left-pointing arrow indicates the upper limit from experiment.}
    \label{fig:fit_result}
\end{figure}

Fig.~\ref{fig:fit_result} shows that Schemes II and III describe all measured partial widths well. In particular, the channels directly affected by rescattering, namely $D^0\bar D^0$, $D^+D^-$, and $D_s^+D_s^-$, are systematically enhanced relative to the short-distance-only predictions. This demonstrates that FSI are quantitatively relevant already at the level of decay rates. The upper limit on $\bar B^0\to D_s^+D_s^-$ is also naturally accommodated.

\begin{figure}[htb!]
    \centering
    \includegraphics[width=0.96\linewidth]{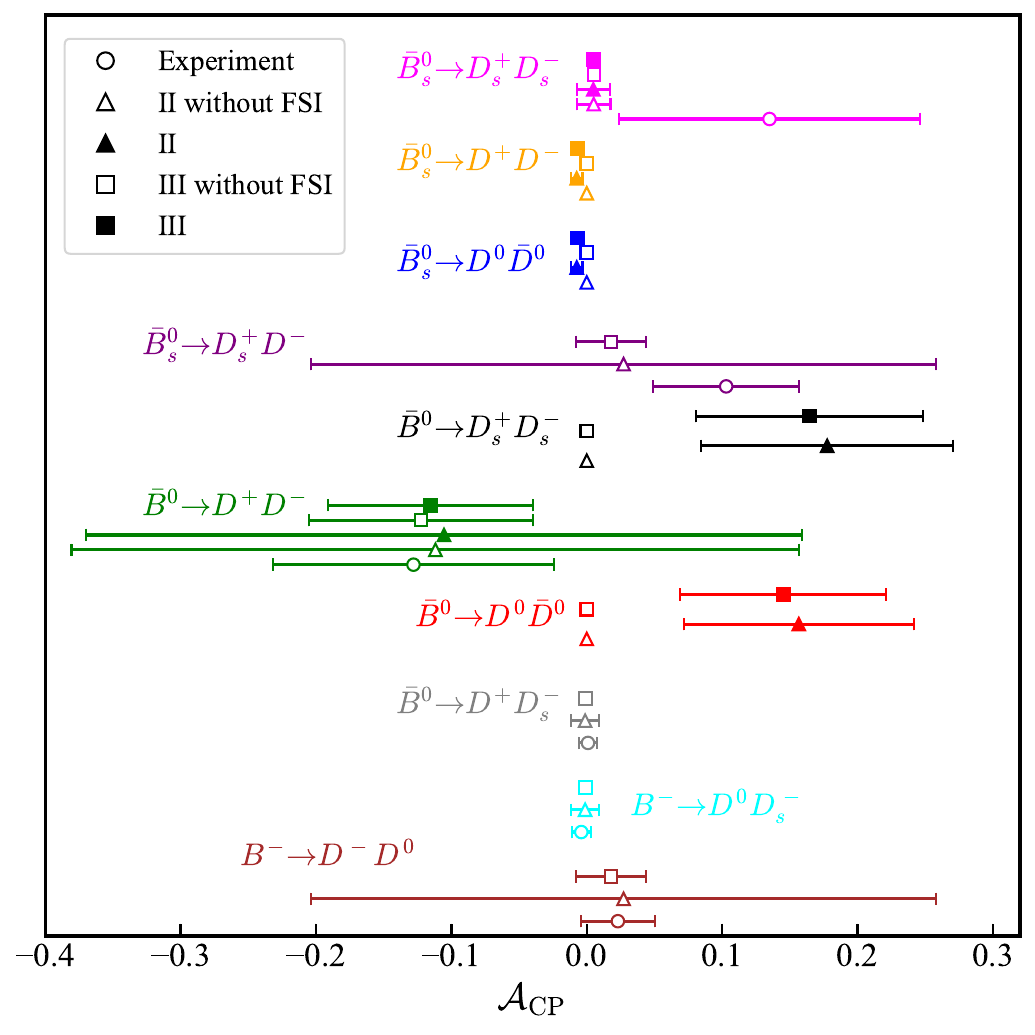}
    \caption{Same notation as in Fig.~\ref{fig:fit_result}, but for the direct CP asymmetries in comparison with available measurements~\cite{LHCb:2018uli,LHCb:2026lvd,LHCb:2024gkk}.}
    \label{fig:acp_all}
\end{figure}

The impact of FSI is even more striking in the CP asymmetries shown in Fig.~\ref{fig:acp_all}. Scheme III yields substantially smaller uncertainties than Scheme II because the measured asymmetries in the FSI-insensitive channels directly constrain $\delta_p$ through Eq.~\eqref{eq:ACPinfit}. Most importantly, rescattering generates sizable asymmetries in the pure-annihilation channels $\bar B^0\to D^0\bar D^0$ and $\bar B^0\to D_s^+D_s^-$, for which any short-distance treatment predicts exactly zero. These channels therefore provide a clean smoking-gun test of long-distance dynamics: a future nonzero measurement of either asymmetry would be direct evidence for rescattering-induced strong phases in hadronic $B$ decays.

For channels such as $\bar B^0\to D^+D^-$, the effect of FSI is not necessarily an enhancement; instead, rescattering can reduce the asymmetry by modifying the interference pattern between short- and long-distance phases. This channel dependence is precisely why a unitary coupled-channel treatment is needed: the same rescattering kernel simultaneously correlates the branching ratios and asymmetries of all channels, leaving no room for channel-by-channel adjustments.

\begin{figure}[htb!]
    \centering
    \includegraphics[width=0.96\linewidth]{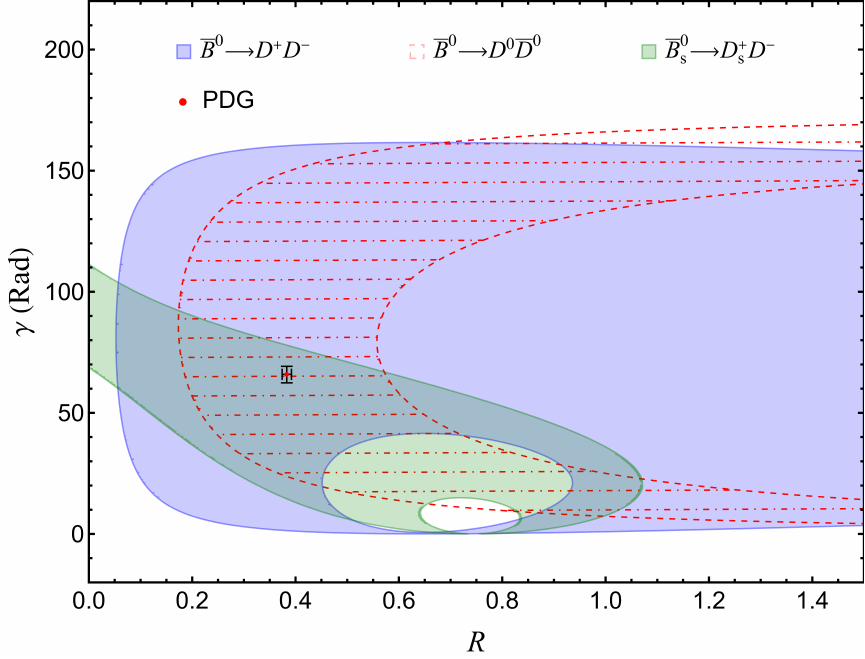}
    \caption{The $\gamma$-$R$ parameter region given by $\mathcal{A}_{\mathrm{CP}}$ of $\bar{B}^0\to D\bar{D}$ and $\bar{B}_s^0\to D_s^+D^-$. The $\bar{B}^0 \to D^+D^-$ channel data restriction determines the blue region, $\bar{B}^0 \to D^0\bar{D}^0$ channel determines the red dashed line region, $\bar{B}_s^0 \to D_s^+D^-$ channel determines the green region, and the red point represents the values from PDG. The $\bar{B}^0 \to D_s^+D_s^-$ and $B^-\to D^0D^-$ channels are omitted -- the former's parameter region coincides with $\bar{B}^0 \to D^0\bar{D}^0$, and the latter's large experimental error renders it unconstraining.}
    \label{fig:parameters_region}
\end{figure}

Finally, the framework has direct implications for CKM phenomenology. Since the amplitudes $T$, $A$, and $P$ are common to all channels, precise measurements of direct CP asymmetries in several modes can be combined to constrain the CKM parameters in a data-driven way. Fig.~\ref{fig:parameters_region} illustrates the resulting $\gamma$-$R$ region from the asymmetries of $\bar B^0\to D^0\bar D^0$, $\bar B^0\to D^+D^-$, and $\bar B_s^0\to D_s^+D^-$. The current uncertainties are still too large for a competitive extraction, but the figure shows that improved measurements in the annihilation-dominated channels would sharply reduce the allowed parameter space. This is another manifestation of the fact that FSI, once controlled, strengthen rather than obstruct CKM studies.

{\it Summary.}~~
We have shown that direct CP asymmetries in $B_{(s)}\to D\bar D$ decays can be described within a predictive final-state-interaction framework in which the rescattering dynamics is fixed independently from $\gamma\gamma\to D\bar D$ data. The all-order solution of the Lippmann-Schwinger equation restores the unitarity that is intrinsically violated in one-loop FSI treatments, and the resulting coupled-channel strong phases provide the missing ingredient needed for a controlled description of CP violation in these decays.

The data strongly favor this picture. A short-distance-only fit fails to reproduce the observed pattern of branching fractions, whereas the inclusion of unitarized FSI improves the fit quality from $\chi^2/{\rm d.o.f}\simeq2.3$ to $\simeq1.0$. At the same time, the fitted short-distance amplitudes remain stable and consistent with independent hadronic estimates, showing that the role of rescattering is not to mimic short-distance physics but to supply the long-distance phases that factorization alone cannot provide.

The most distinctive consequence is the emergence of sizable direct CP asymmetries in the pure-annihilation channels $\bar B^0\to D^0\bar D^0$ and $\bar B^0\to D_s^+D_s^-$. Since these asymmetries vanish identically in any short-distance treatment, a nonzero measurement in either channel would constitute a direct experimental signature of long-distance rescattering dynamics. This makes these modes especially powerful null tests of the short-distance picture and clean discovery channels for FSI effects.

Our results therefore establish final-state interactions not as an irreducible source of hadronic uncertainty, but as a falsifiable and data-driven ingredient of the Standard Model. Future measurements of the yet-unmeasured branching fractions and direct CP asymmetries at LHCb and Belle II can directly test this framework, with immediate implications for CKM parameter extractions and for the interpretation of possible new-physics signals in nonleptonic $B$ decays.

\vspace{0.2cm}
{\bf \color{gray}Acknowledgements:}~~We are grateful to Feng-Kun Guo and Rui-Hui Li for their useful discussions.
This work is partly supported by the National Natural Science Foundation of China with Grants Nos.~12375073, ~12547105, 12125503 and 12305103. 

\clearpage
\bibliography{ref.bib}

\onecolumngrid
\clearpage
\newpage
\appendix

{\large \bf \section*{Supplemental Materials}}

\section{Topology Diagrams}
\label{app:Topology Diagrams}
\begin{figure}[h!]
\centering
    \subfigure[DE: Disconnected Emission]{
    \label{fig:toplogic:DE}
    \includegraphics[width=0.375\linewidth]{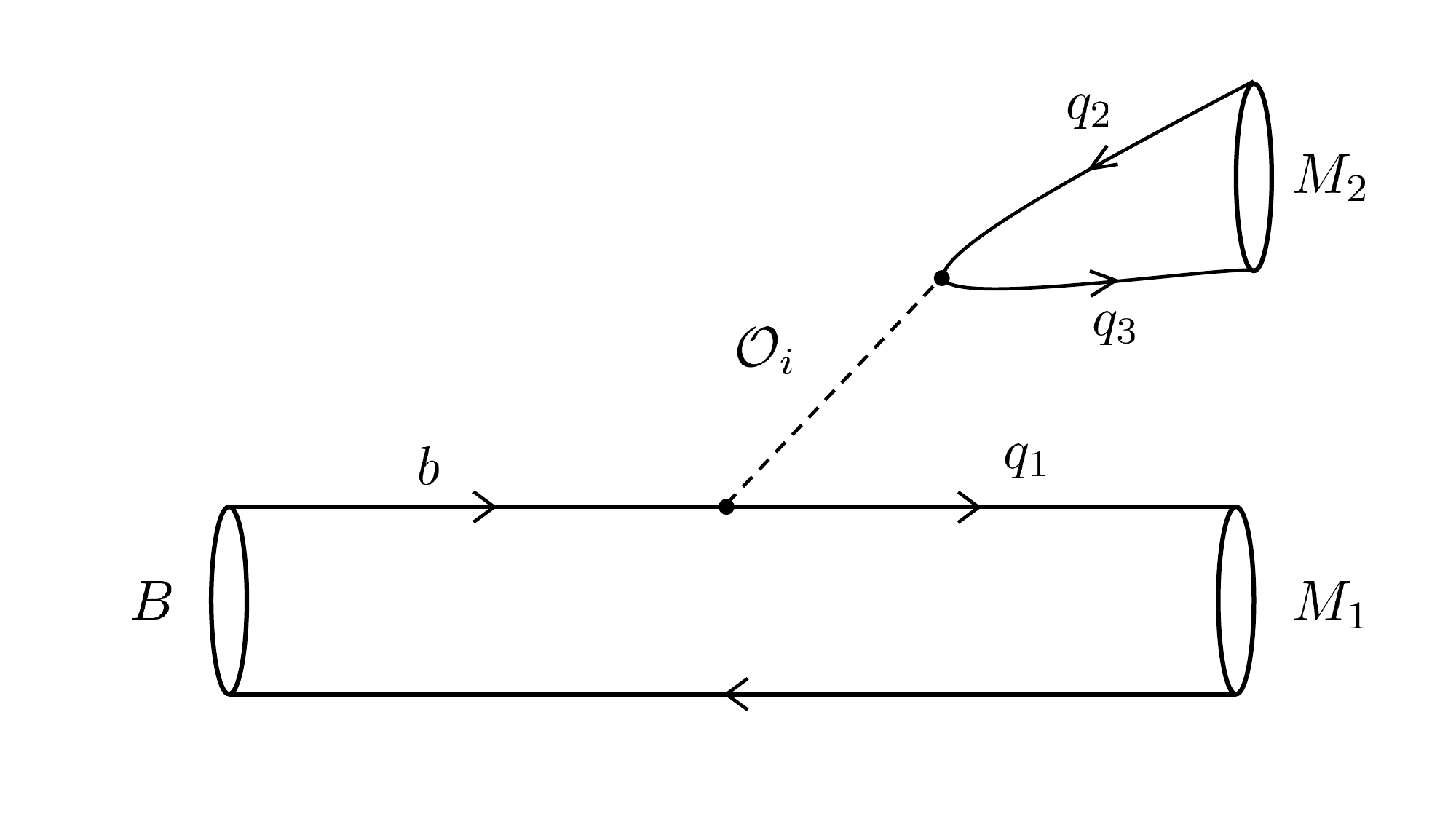}
    }
    \subfigure[CE: Connected Emission]{
    \label{fig:toplogic:CE}
    \includegraphics[width=0.375\linewidth]{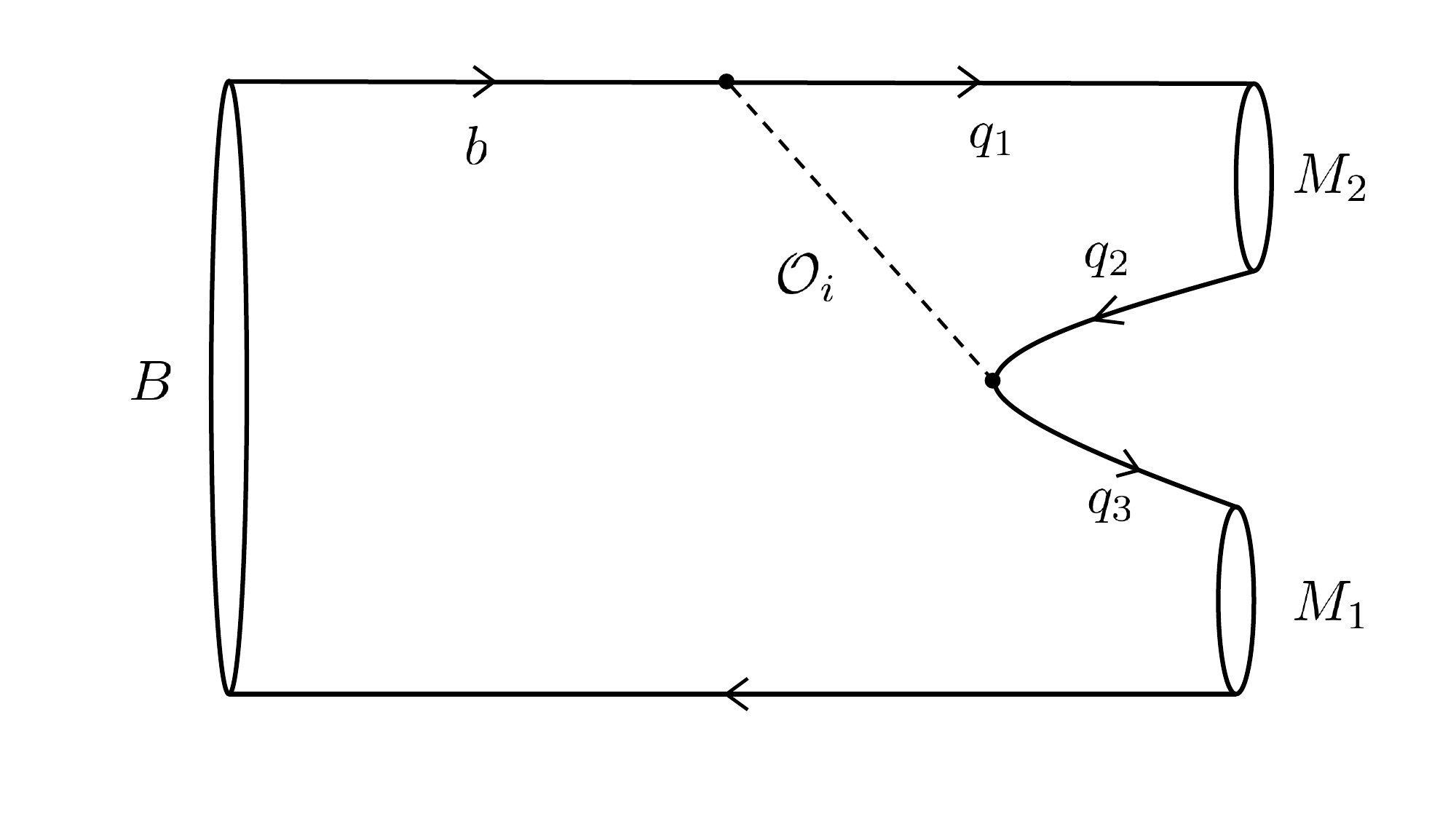}
    }
    \subfigure[DA: Disconnected Annihilation]{
    \label{fig:toplogic:DA}
    \includegraphics[width=0.375\linewidth]{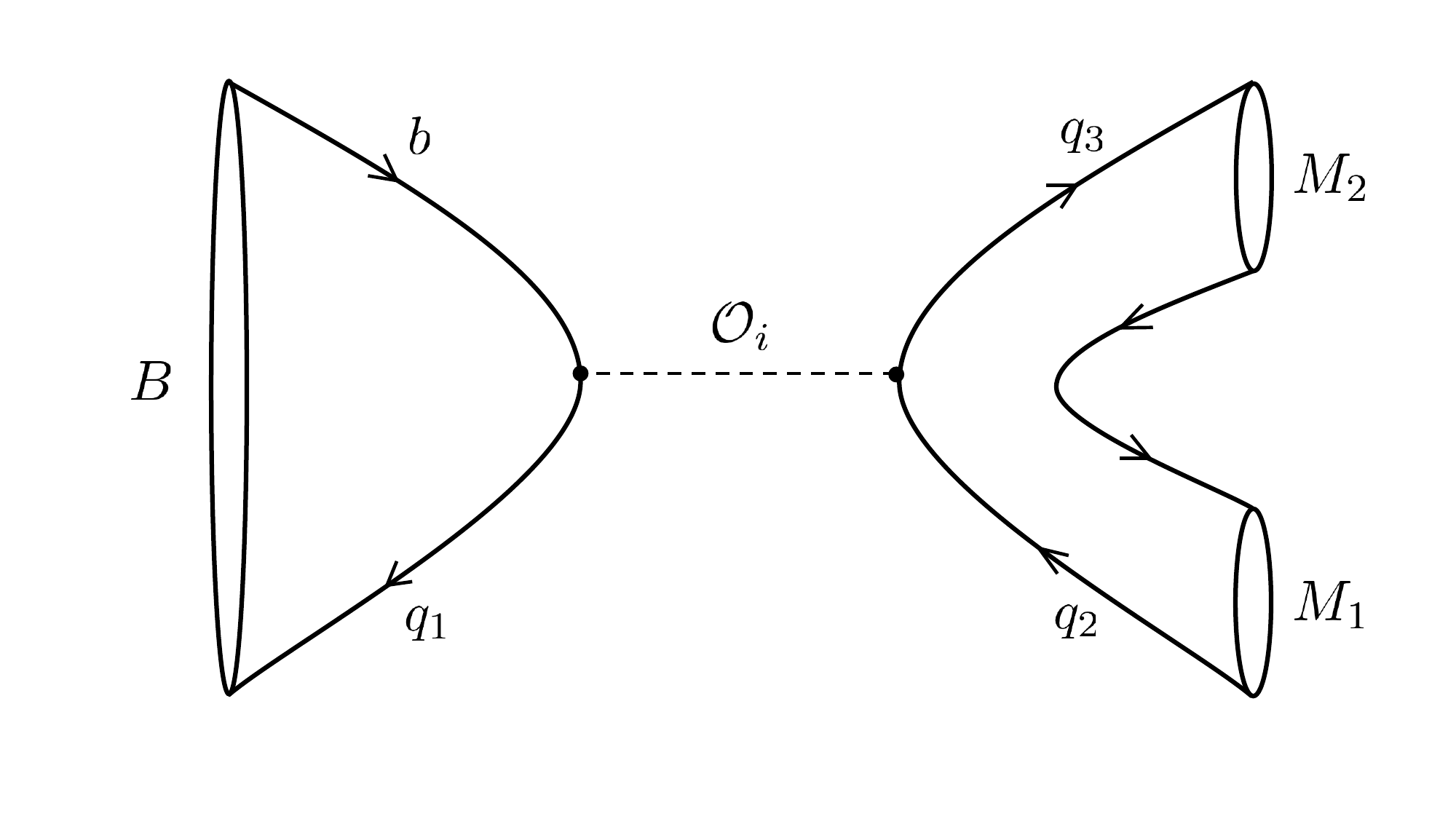}
    }
    \subfigure[CA: Connected Annihilation]{
    \label{fig:toplogic:CA}
    \includegraphics[width=0.375\linewidth]{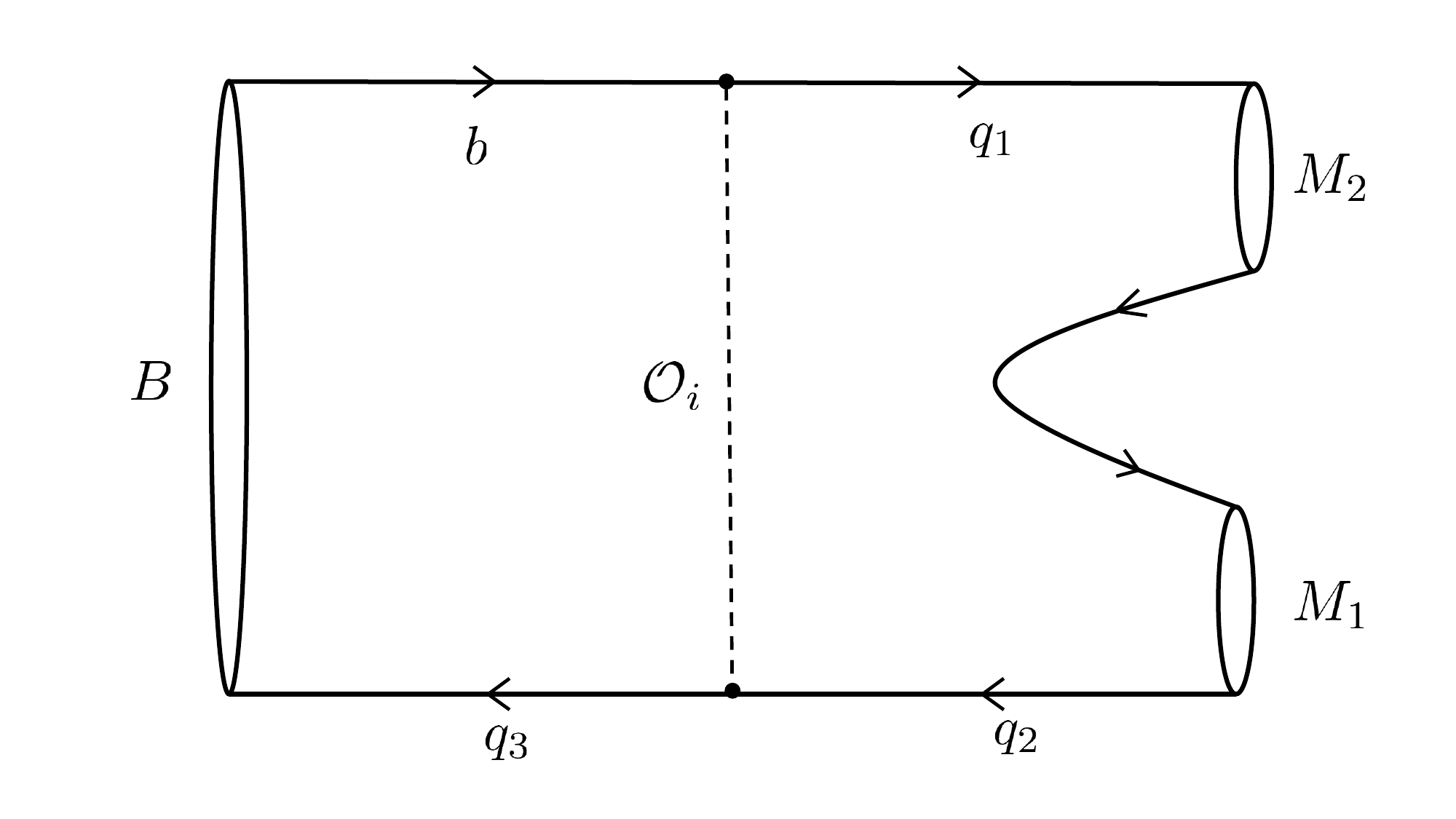}
    }
    \subfigure[DP: Disconnected Penguin]{
    \label{fig:toplogic:DP}
    \includegraphics[width=0.375\linewidth]{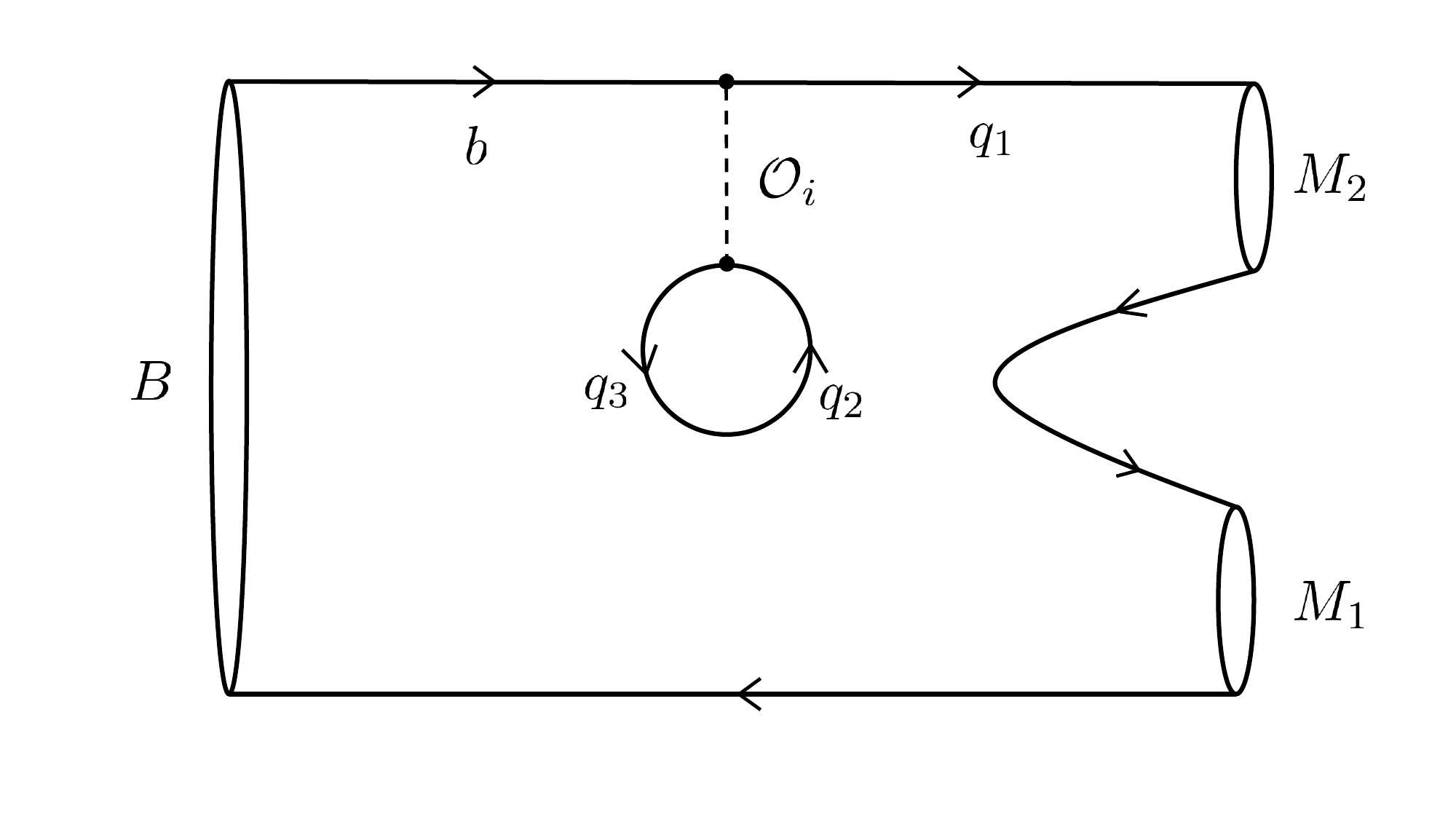}
    }
    \subfigure[CP: Connected Penguin]{
    \label{fig:toplogic:CP}
    \includegraphics[width=0.375\linewidth]{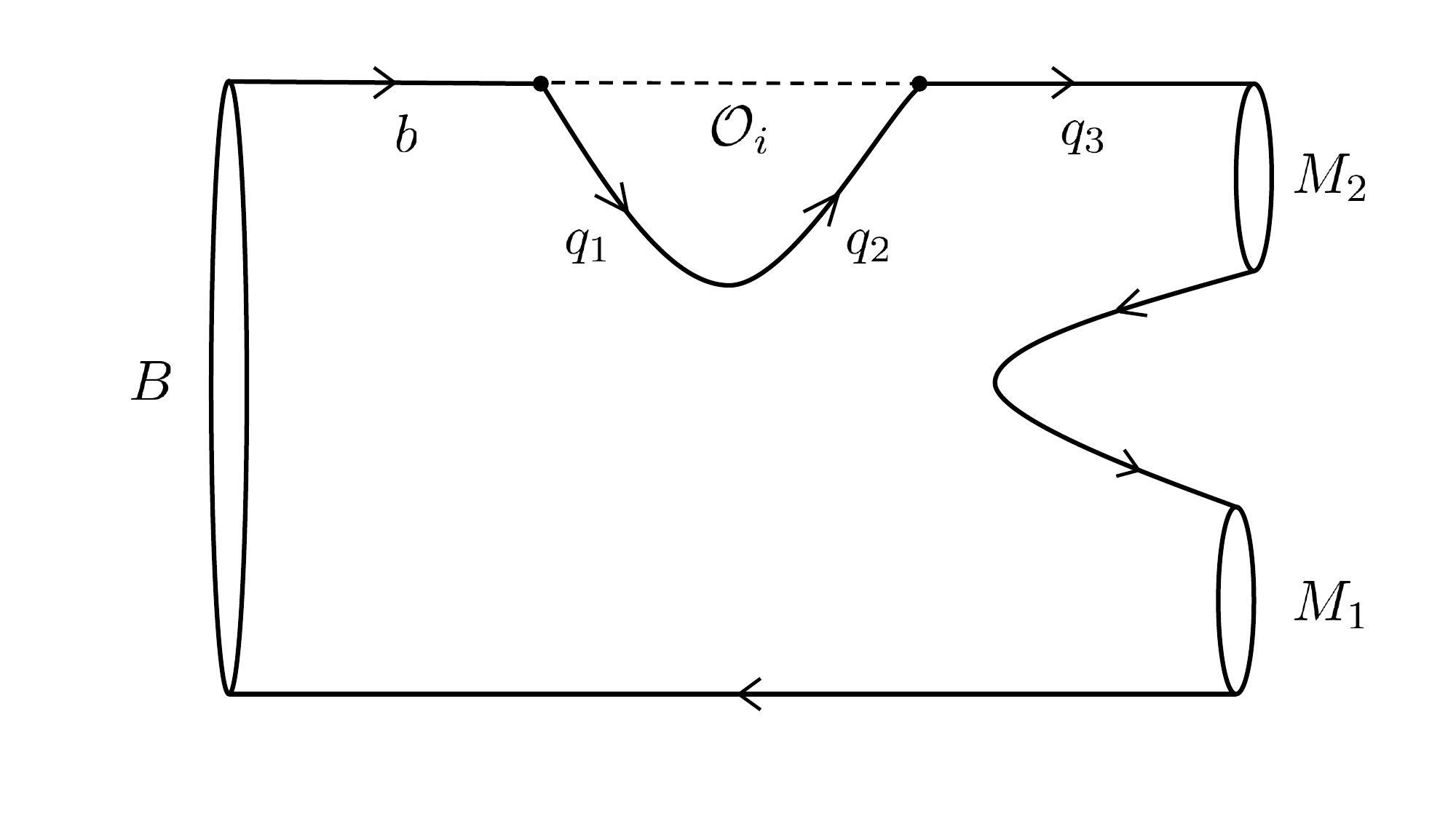}
    }
    \subfigure[DPA: Disconnected Penguin-Annihilation]{
    \label{fig:toplogic:DPA}
    \includegraphics[width=0.375\linewidth]{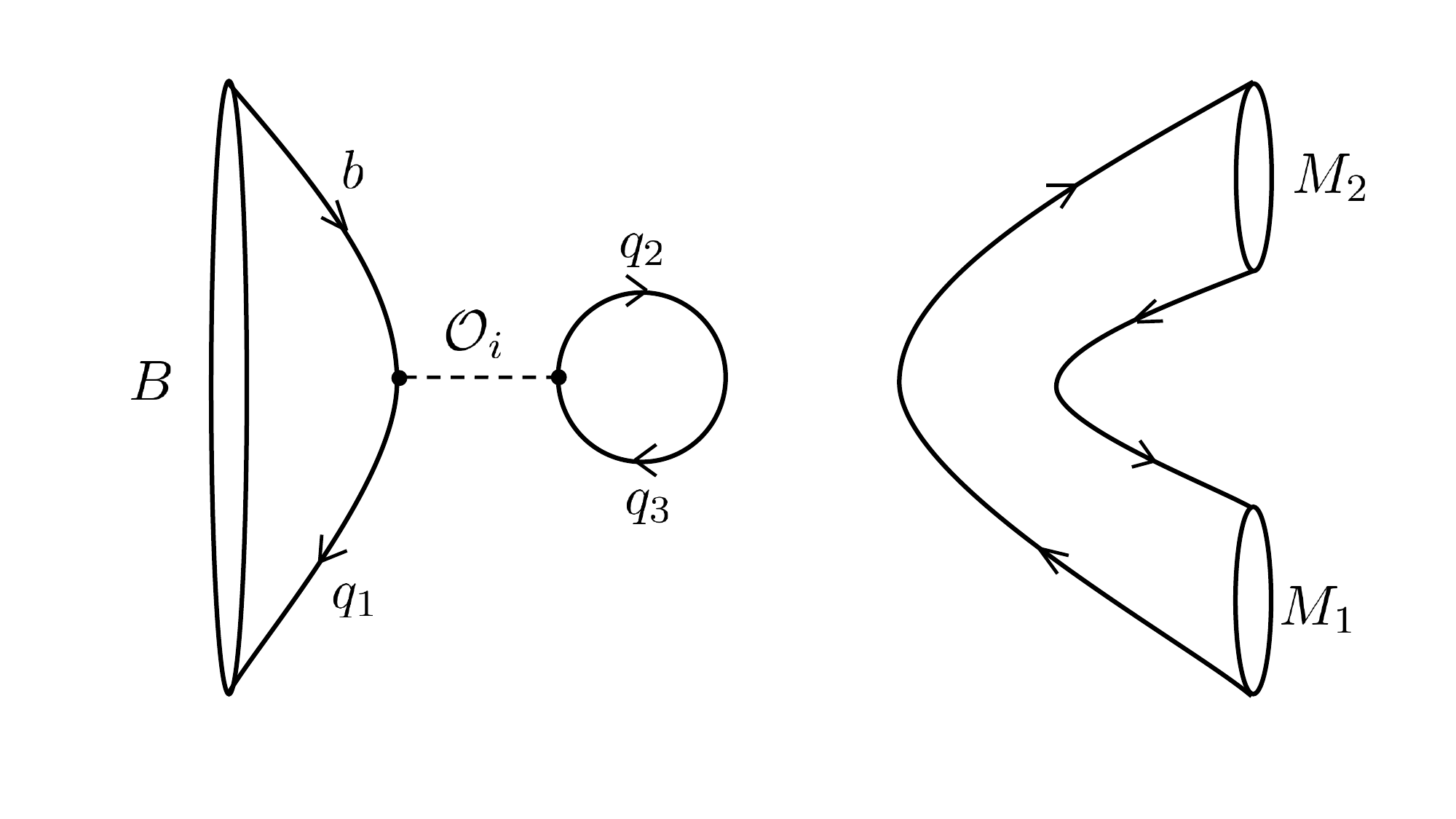}
    }
    \subfigure[CPA: Connected Penguin-Annihilation]{
    \label{fig:toplogic:CPA}
    \includegraphics[width=0.375\linewidth]{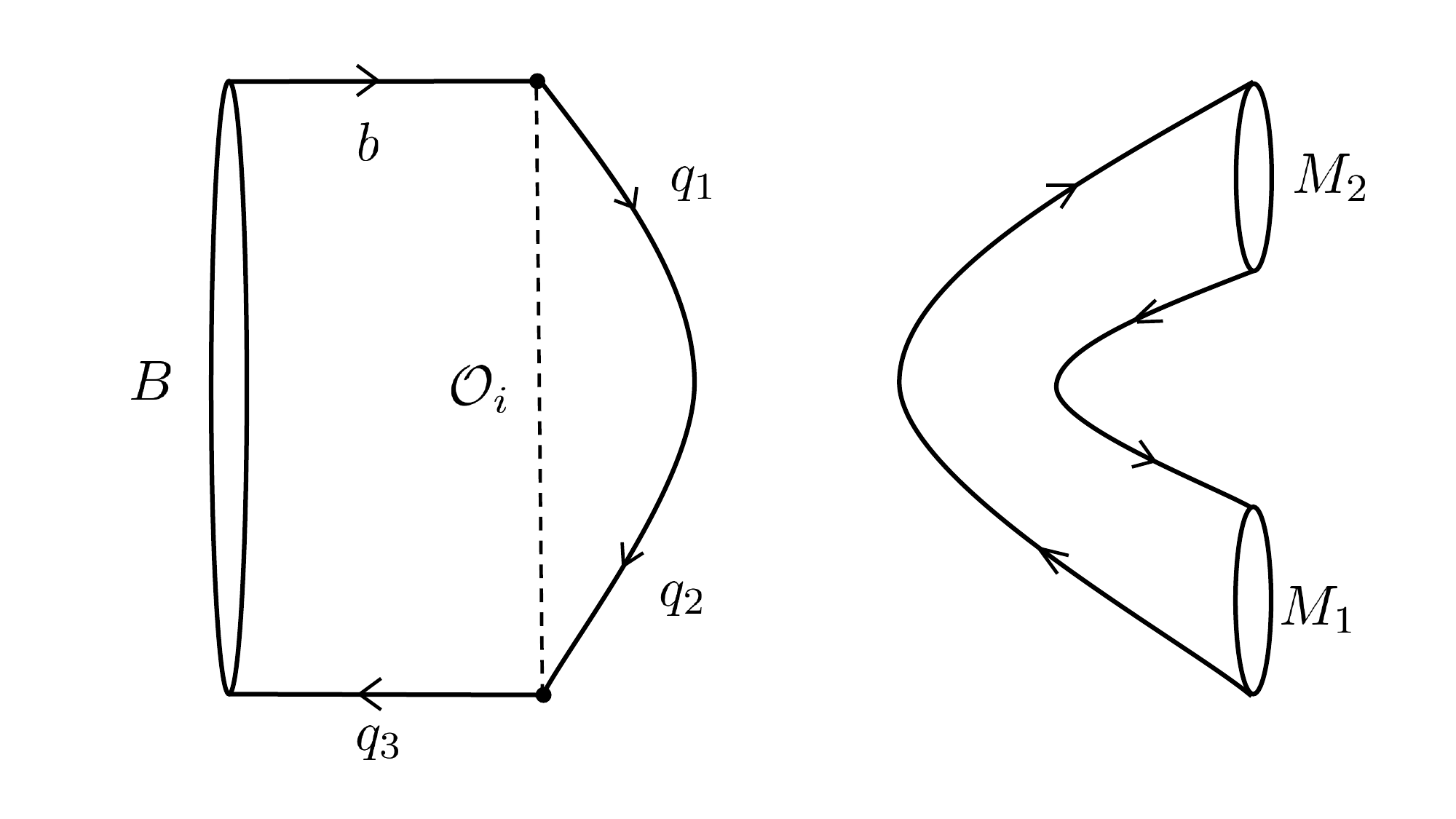}
    }
    \caption{The topologies considered in this work.}
    \label{fig:toplogic}
\end{figure}

The current-current operators, $\mathcal{O}_{1}$ and $\mathcal{O}_{2}$, have the same chiral structure, but different color structures, which will be reflected in corresponding Wilson coefficients. For $B\to F$ decays, the long-distance effects (energy scale below $\mu=m_b$) are contained in the hadronic matrix elements $\left\langle F\right |\mathcal{O}_{i}(\mu)\left| B\right\rangle$. These hadronic matrix elements contain various contributions and can be distinguished by their Wick contractions~\cite{Buras:1998ra} as shown in Fig.~\ref{fig:toplogic}.
In total, there are eight topologies in $B\to D\bar{D}$ processes. 
The short-range effects (energy scale above $\mu=m_b$) of such decays are contained in the Wilson coefficients corresponding to $\mathcal{O}_i$, which can be perturbatively calculated by QCD.
Varying topologies arising from the operators $\mathcal{O}_i$	
derive different effective parameters and are listed below, with the Wilson coefficients absorbed into the parameter.
\begin{itemize}
    \item Emission topologies
\begin{equation}
    E^{c}=C_1(\mu)\left\langle \mathcal{O}_1(\mu) \right \rangle_{\mathrm{CE}}^{c}+C_2(\mu)\left\langle \mathcal{O}_2(\mu) \right \rangle_{\mathrm{DE}}^{c}.
\end{equation}
    \item Annihilation topologies
\begin{equation}
    A^{c}=C_1(\mu)\left\langle \mathcal{O}_1(\mu) \right \rangle_{\mathrm{DA}}^{c}+C_2(\mu)\left\langle \mathcal{O}_2(\mu) \right \rangle_{\mathrm{CA}}^{c}.
\end{equation}
    \item Penguin topologies
\begin{equation}
    P^{q}=C_1(\mu)\left\langle \mathcal{O}_1(\mu) \right \rangle_{\mathrm{DP}}^{q}+C_2(\mu)\left\langle \mathcal{O}_2(\mu) \right \rangle_{\mathrm{CP}}^{q}.
\end{equation}
    \item Penguin Annihilation topologies \begin{equation}
    PA^{q}=C_1(\mu)\left\langle \mathcal{O}_1(\mu) \right \rangle_{\mathrm{DPA}}^{q}+C_2(\mu)\left\langle \mathcal{O}_2(\mu) \right \rangle_{\mathrm{CPA}}^{q}.
\end{equation}
\end{itemize}
Here, $\left\langle \mathcal{O}_{i}(\mu) \right \rangle^{q}$ denotes a current-current operator $\mathcal{O}i^{qq^{\prime}}(\mu)$ inserted into a given topology diagram. Various topologies are distinguished by their subscripts. For example, $\left\langle \mathcal{O}_1(\mu) \right \rangle_{\mathrm{CE}}^{c}$ represents the operator $\mathcal{O}_{1}^{cq^{\prime}}(\mu)$ inserted into the CE topology. When neglecting CKM-suppressed contributions, i.e. $A^u$ and $PA^{u}$, 
the decay amplitudes depend on three independent parameters 
\begin{equation}
T\equiv E^{c}+P^{c}, \quad A\equiv A^{c}+PA^{c}, \quad P\equiv P^{u}.
\end{equation}
Finally, the explicit weak decay amplitudes for each channel are given  in Tab.~\ref{tab:decay channel}.

\begin{table*}[htpb!]
    \centering
    \caption{The weak decay amplitudes of bottomed meson into a pair of charmed pseudo-scalar mesons. $Y_{qq^\prime}
    \equiv V_{qb}^* V_{qq^\prime}$ is a product of the CKM matrix elements.
    }
    \renewcommand{\arraystretch}{1.5}
    \resizebox{0.8\textwidth}{!}{
    \begin{tabular}{l@{\hspace{25pt}}c@{\hspace{40pt}}l@{\hspace{25pt}}c}
       \hline
        Channel& Amplitude& Channel & Amplitude\\
        \hline
       $B^-\to D^0D^-$  & $Y_{cd}T+Y_{ud}P$ & $\bar{B}^0 \to D_s^+D_s^-$&$Y_{cd}A$\\
        $B^-\to D^0D_s^-$& $Y_{cs}T+Y_{us}P$ & $\bar{B}_s^0\to D^+D^-$& $Y_{cs}A$\\
        $\bar{B}^0 \to D^+ D^-$& $Y_{cd}(T+A)+Y_{ud}P$ &$\bar{B}_s^0\to D_s^+D_s^-$ & $Y_{cs}(T+A)+Y_{us}P$\\
        $\bar{B}^0 \to D^+D_s^-$& $Y_{cs}T+Y_{us}P$ &$\bar{B}_s^0\to D^0\bar{D}^0$&$Y_{cs}A$\\
        $\bar{B}^0 \to D^0\bar{D}^0$&$Y_{cd}A$ & $\bar{B}_s^0 \to D_s^{+}D^-$&$Y_{cd}T+Y_{ud}P$\\
        \hline
    \end{tabular}
    }
    \label{tab:decay channel}
\end{table*}

\section{The details of FSI}
\label{app:FSI}

Since the charm quark mass $m_c\sim 1.5~\mathrm{GeV}$ is much larger than the $\Lambda_{\mathrm{QCD}}\sim 0.3~\mathrm{GeV}$, we would expect that heavy quark spin symmetry(HQSS) approximately works well in the $D\bar{D}$ system. The leading order(LO) Lagrangian respecting HQSS reads as~\cite{Wise:1992hn,AlFiky:2005jd,Ji:2022uie}
\begin{equation}
    \begin{aligned}
        \mathcal{L}_{4H}&=\frac{1}{4}\mathrm{Tr}[\bar{H}^{(Q)a}H_b^{(Q)}\gamma_{\mu}]\mathrm{Tr}[H^{(\bar{Q})c}\bar{H}_d^{(\bar{Q})}\gamma^{\mu}]\\
        &\times(F_{A}\delta_{a}^{b}\delta_{c}^{d}+F_{A}^{\lambda}\vec{\lambda}_{a}^{b}\cdot\vec{\lambda}_{c}^{d})+\frac{1}{4}\mathrm{Tr}[\bar{H}^{(Q)a}H_b^{(Q)}\gamma_{\mu}\gamma_5]\\
        &\times\mathrm{Tr}[H^{(\bar{Q})c}\bar{H}_d^{(\bar{Q})}\gamma^{\mu}\gamma_5](F_{B}\delta_{a}^{b}\delta_{c}^{d}+F_{B}^{\lambda}\vec{\lambda}_{a}^{b}\cdot\vec{\lambda}_{c}^{d}),
    \end{aligned}
\end{equation}
where $F_{A(B)}^{(\lambda)}$ are the low-energy constants, and $\vec{\lambda}$ denote the eight Gell-Mann matrices in the $\mathrm{SU}(3)_\mathrm{f}$ space. The super field $H_a^{(Q)}$($H^{(\bar{Q})}_a$)
\begin{equation}
    \begin{aligned}
        &H_a^{(Q)}=\frac{1+\mathbf{\cancel{v}}}{2}(V^{(Q)}_{a\mu}\gamma^{\mu}-P^{(Q)}_a\gamma_5),\quad v\cdot V_a^{(Q)}=0,\\
        &H^{(\bar{Q})}_a=(V^{(\bar{Q})}_{a\mu}\gamma^{\mu}-P^{(\bar{Q})}_a\gamma_5)\frac{1-\mathbf{\cancel{v}}}{2},\quad v\cdot V_a^{(\bar{Q})}=0,
    \end{aligned}
    \end{equation}
    annihilates a $P_a^{(Q)}$($P_a^{(\bar{Q})}$) and a $V_{a\mu}^{(Q)}$($V_{a\mu}^{(\bar{Q})}$) particle, and $v$ is the velocity of the particle.
Here $P_a^{(Q)}$ denote a pseudo-scalar charmed meson annihilation operator with $a=1,2,3$ denoting $D^0,D^+,D_s^+$, respectively. $V_{a\mu}^{(Q)}=(D^{*0}_{\mu},D^{*+}_{\mu},D_{s\mu}^{*+})$ denote vector charmed meson annihilation operators. The $P_a^{(\bar{Q})}=CP_a^{Q}C^{-1}$ and $V_{a\mu}^{(\bar{Q})}=-CV_{a\mu}^{(Q)}C^{-1}$ represent the corresponding anti-charmed meson annihilation operators, respectively, where $C$ is the charge conjugation transform. In this case, for the $D\bar{D}$ system, the Lagrangian can be simplified as
\begin{equation}
    \mathcal{L}_{4H}=v^2P^{(Q)a\dagger}P_b^{(Q)}P^{(\bar{Q})c}P_d^{(\bar{Q})\dagger}(F_{A}\delta_{a}^{b}\delta_{c}^{d}+F_{A}^{\lambda}\vec{\lambda}_{a}^{b}\cdot\vec{\lambda}_{c}^{d})
\end{equation}
with $v^2=1$ in the heavy quark limit.
Taking the non-relativistic (NR) approximation, the $D\bar{D}$ contact potential reads as~\footnotemark
\footnotetext{Although other works have suggested the possible existence of a $\chi_{c0}(2P)$~\cite{Guo:2012tv} state around $3.9~\mathrm{GeV}$, the $D\bar{D}$ dynamics is purely determined by the $\gamma\gamma\to D\bar{D}$ cross sections below $4.3~\mathrm{GeV}$. 
Moreover, we have verified that the inclusion of a bare charmonium state does not even affect the dynamics at the 
$B$ meson mass.}
\begin{equation}
    V=\begin{pmatrix}
        F_{A}+\frac{4}{3}F_{A}^{\lambda} & 2F_{A}^{\lambda} & 2F_{A}^{\lambda}\\
        2F_{A}^{\lambda} & F_{A}+\frac{4}{3}F_{A}^{\lambda} & 2F_{A}^{\lambda}\\
        2F_{A}^{\lambda} & 2F_{A}^{\lambda} & F_{A}+\frac{4}{3}F_{A}^{\lambda}
    \end{pmatrix}.
\end{equation}

In contrast to one-loop treatments that iterate a single rescattering, the coupled-channel $T$-matrix is solved non-perturbatively by using the Lippmann-Schwinger equation,
\begin{equation}
    t(s)=V+VG^{\Lambda}(s)t(s).
    \label{eq:LS equation}
\end{equation}
where $s=(p_1+p_2)^2$, $t=(p_1-p_3)^2$, $u=(p_1-p_4)^2$ are Mandelstam variables, with $p_{1,3}$ and $p_{2,4}$ the four-momenta of the initial and final particles, respectively. $G^{\Lambda}(s)$ is a diagonal matrix in which the $i$-th nonzero element is given by
\begin{equation}
\begin{aligned}
    G_i^{\Lambda}(s)&=-i\int\frac{d^4q}{(2\pi)^4}\frac{f^{\Lambda}(\mathbf{q}^2)}{(q^2-m_i^2+i\epsilon)[(P-q)^2-m_i^2+i\epsilon]}\\
    &=-\frac{1}{4m_i}\frac{1}{2\pi^2}\int_{0}^{\infty} d|\mathbf{q}|~\frac{\mathbf{q}^2}{k_i^2-\mathbf{q}^2+i\epsilon}e^{-\frac{2\mathbf{q}^2}{\Lambda^2}}\\
        &=-\frac{1}{4m_i}\frac{1}{2\pi^2}\left(\mathcal{P}\int_0^{\infty}d|\mathbf{q}|\frac{\mathbf{q}^2}{k_i^2-\mathbf{q}^2}e^{-\frac{2q_i^2}{\Lambda^2}}-i\pi\frac{k_i}{2}e^{-\frac{2k_i^2}{\Lambda^2}}\right)\\
        &=-\frac{1}{4m_i}\frac{1}{2\pi^2}\left(-\int_0^\infty  d|\mathbf{q}|~e^{-\frac{2q_i^2}{\Lambda^2}}+k_i^2e^{-\frac{2k_i^2}{\Lambda^2}}\mathcal{P}\int_0^{\infty}d|\mathbf{q}|\frac{e^{\frac{2}{\Lambda^2}(k_i^2-q_i^2)}}{k_i^2-\mathbf{q}^2}-i\pi\frac{k_i}{2}e^{-\frac{2k_i^2}{\Lambda^2}}\right)\\
        &=-\frac{1}{4m_i}\frac{1}{2\pi^2}\left(-\frac{\Lambda \sqrt{\pi}}{2\sqrt{2}}+k_i^2e^{-\frac{2k_i^2}{\Lambda^2}}\frac{\pi}{2k_i}\mathrm{erfi}(\frac{\sqrt{2}k_i}{\Lambda})-i\pi\frac{k_i}{2}e^{-\frac{2k_i^2}{\Lambda^2}}\right)\\
        &=\frac{1}{16\pi m_i}\left[\frac{\Lambda}{\sqrt{2\pi}}-k_{i}e^{-\frac{2k_i^2}{\Lambda^2}}\left(\mathrm{erfi}(\frac{\sqrt{2}k_i}{\Lambda})-i\right)\right],
        \end{aligned}
    \label{eq:NR two point correction function}
\end{equation}
where $P=p_1+p_2=p_3+p_4$ is the total four momentum. $f^{\Lambda}(\mathbf{q}^2)=\mathrm{exp}(-2\mathbf{q^2}/\Lambda^2)$ is a Gaussian form factor with $\Lambda$ a parameter, which is used to suppress the high-energy contribution in the integral and enters in each vertex. 
where $\mathrm{erfi}(z)$ is the standard imaginary error function and denoted by
\begin{equation}
    \mathrm{erfi}(z)=\frac{2}{\sqrt{\pi}}\int_0^{z}dt~e^{t^2}.
\end{equation}
The above two-body propagator $G^{\Lambda}_i(s)$ breaks unitarity, as $\mathrm{Im}G_{i}^{\Lambda}(s)\propto k_{i}\mathrm{exp}(-\frac{2k_i^2}{\Lambda^2})$ as seen from Eq.\eqref{eq:NR two point correction function}. To restore unitarity, the same form factor is also  introduced to all external particles.

In  Eq.~\eqref{eq:NR two point correction function}, the overall factor $\frac{1}{4m_i^2}$ can be absorbed into the normalization of the non-relativistic initial and final state wave functions.
As the linear term in Eq.~\eqref{eq:NR two point correction function} makes the parameters $\Lambda$ largely correlated with the parameters in the potential, we neglect this term. As an exponential form factor is introduced in the two-body propagator, it should also appear in each vertex of the potential. 

\begin{figure*}[htp!]
    \centering
    \includegraphics[width=0.98\textwidth]{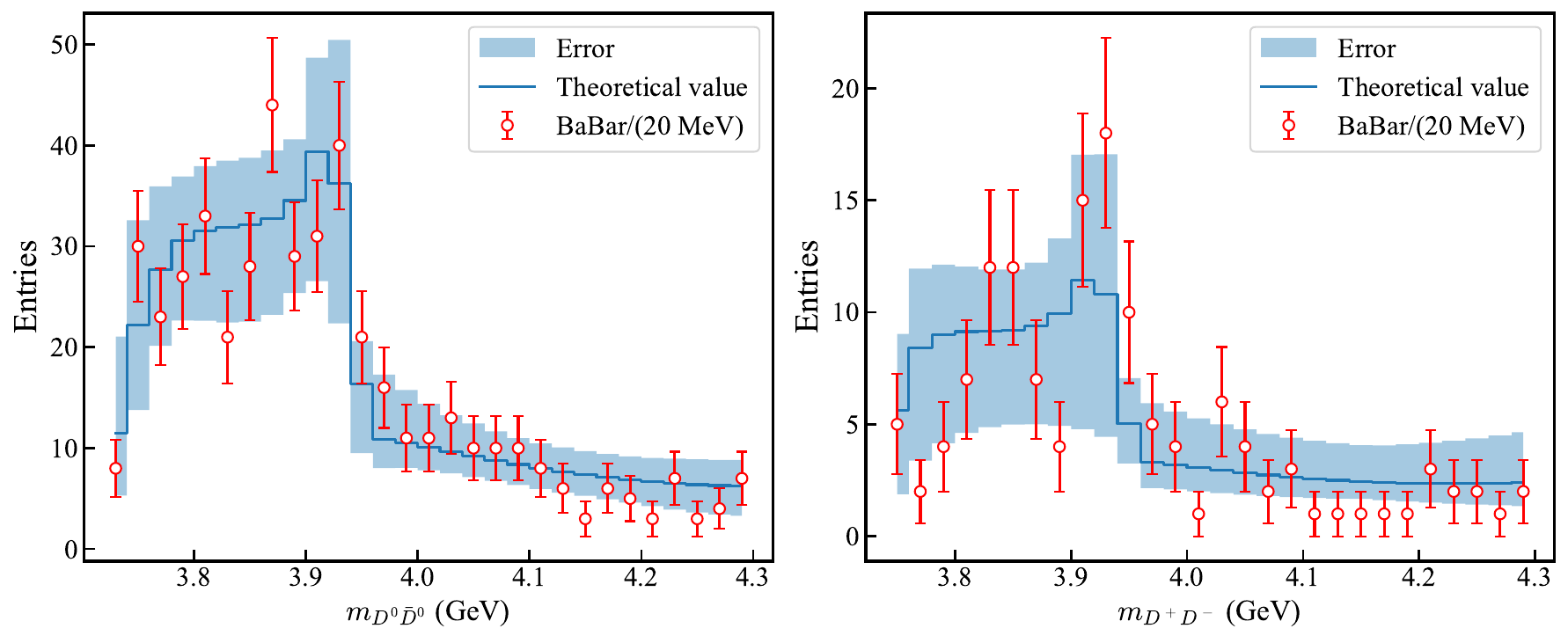}
    \caption{From left to right, the  $\gamma\gamma\to D^0\bar{D}^0$ and $\gamma\gamma\to D^+D^-$ event distributions, respectively. Experimental values come from Ref.~\cite{BaBar:2010jfn}}
    \label{fig:fit_FSI}
\end{figure*}

\begin{figure}[htp!]
    \centering
    \includegraphics[width=0.98\textwidth]{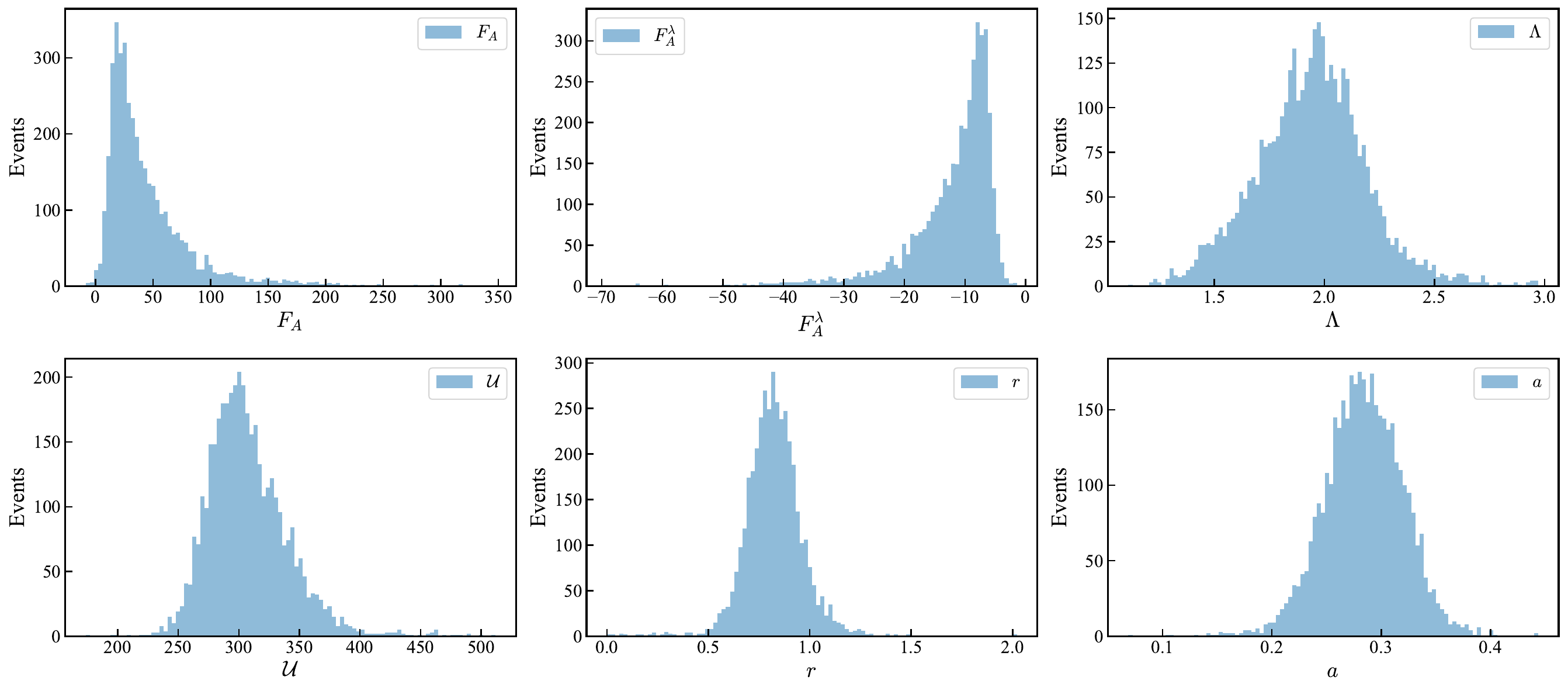}
    \caption{The distribution of the parameters approximates a normal distribution. The average of the distribution represents the central values of the parameters, and the standard deviation represent the errors.}
    \label{fig:para_dis}
\end{figure}

In this work, we only consider the $S$-wave $D\bar{D}$ interactions in the $\gamma\gamma\to D\bar{D}$ process, with the amplitude of the $i$-th channel reads as 
\begin{equation}
    \mathcal{M}(\gamma\gamma\to D^i\bar{D}^i)=\mathcal{C}_jG^{\Lambda}_j(s)t_{ji}(s)+\mathcal{C}_i,
\end{equation}
where $\mathcal{C}_i$ is the bare production amplitude for the $i$-th channel, with $i=1,2,3$ for the $D^0\bar{D}^0$, $D^+D^-$, $D_s^+D_s^-$ channels, respectively. Here the bare production amplitudes are parameterized as $\mathcal{C}_1=r\mathrm{U}$, $\mathcal{C}_2=\mathrm{U}$, $\mathcal{C}_3=\mathrm{U}$, with $\mathcal{U}$ the bare production amplitude for  $\gamma\gamma$ directly create $D^+D^-$ or $D_s^+D_s^-$, and $r$ the relative strength between the neutral channel and the charged one. $G^{\Lambda}_{j}(s)$ represent the two-body propagator for $j-$th channel, and $t_{ij}(s)$ represent the $T-$matrix element of $D\bar{D}$ scattering.
To render the contribution from the high-energy region, a Gaussian form factor is introduced to the two-body propagator. Then Eq.\eqref{eq:LS equation} becomes 
\begin{equation}
    F^{\Lambda}(s)t(s)F^{\Lambda}(s)=F^{\Lambda}(s)VF^{\Lambda}(s)+F^{\Lambda}(s)VG^{\Lambda}(s)t(s)F^{\Lambda}(s),
\end{equation}
where $F^{\Lambda}(s)$ is a diagonal matrix in which the ith nonzero element is given by
$F_{i}^{\Lambda}(s)=\mathrm{exp}(-k_i^2/\Lambda^2)$.
Consequently,
\begin{equation}
    t(s)=F^{\Lambda}(s)[1 -VG^{\Lambda}(s)]^{-1}VF^{\Lambda}(s),
\end{equation}
where we have replaced $F^{\Lambda}(s)t(s)F^{\Lambda}(s)$ by $ t(s)$. Finally, the unitarity of $S$-matrix 
\begin{equation}
    \mathrm{Im}t(s)=t(s)\Sigma(s)t^{\dagger}(s)
    \label{eq:T unitarity}
\end{equation}
is preserved~\cite{GomezNicola:2001as}, 
where $\Sigma(s)$ is a diagonal matrix in which the $i$-th nonzero element is given by $\Sigma_{i}(s)=k_{i}/\sqrt{s}$.
Finally, we derive the scattering cross section for $\gamma\gamma\to D\bar{D}$ reaction,
\begin{equation}
    \sigma_i(s)=\frac{\sqrt{s(s-4m_{i}^2)}}{16\pi s^2}|\mathcal{M}(\gamma\gamma\to D^i\bar{D}^i)|^2.
\end{equation}
In order to consider experimental resolution of the $\gamma\gamma\to D\bar{D}$ process, we convolute a Gaussian function to the theoretical $\gamma\gamma\to D\bar{D}$ cross sections. The event of the $n$-th bin in the $i$-th channel reads as~\cite{Zhang:2022hfa}
\begin{equation}
    \begin{aligned}
        N_i(n)&=\frac{a_i}{E_{n+1}-E_n}\int_{E_{n}}^{E_{n+1}}dE\int_{-\infty}^{\infty}dE^{\prime}~\sigma[(E-E^{\prime})^2]\frac{1}{\sqrt{2\pi}\sigma}\mathrm{exp}(-\frac{E^{\prime2}}{2\sigma^2})\\
        &=\frac{a_i}{E_{n+1}-E_n}\int_{E_n}^{E_{n+1}}dE\int_{-\infty}^{\infty}dE^{\prime}~\sigma(E^{\prime2})\frac{1}{\sqrt{2\pi}\sigma}\mathrm{exp}(-\frac{(E-E^{\prime})^2}{2\sigma^2})\\
        &=\frac{a_i}{E_{n+1}-E_{n}}\int_{-\infty}^{\infty}dE^{\prime}~\sigma(E^{\prime2})\int_{E_n}^{E_{n+1}}dE~\frac{1}{\sqrt{2\pi}\sigma}\mathrm{exp}(-\frac{(E-E^{\prime})^2}{2\sigma^2})\\
        &=\frac{a_i}{2(E_{n+1}-E_n)}\int_{-\infty}^{\infty}dE^{\prime}\ \sigma(E^{\prime2})\left[\mathrm{Erf}\left(\frac{E_{n}-E^{\prime}}{\sqrt{2}\sigma}\right)-\mathrm{Erf}\left(\frac{E_{n+1}-E^{\prime}}{\sqrt{2}\sigma}\right)\right],
    \end{aligned}
\end{equation}
where $E_{n+1}$ and $E_{n}$ are the upper and lower for $n$-th bin, and $E_{n+1}-E_{n}\equiv 0.02~\mathrm{GeV}$ is the bin size of the experimental data. $\sigma=0.01~\mathrm{GeV}$ is the experimental resolution. We set  the integration range from $\frac{E_{n+1}+E_n}{2}-3\sigma$ to $\frac{E_{n+1}+E_n}{2}+3\sigma$, instead of from  negative infinity to positive infinity,
\begin{equation}
    N_i(n)=\frac{a_i}{2\times0.02}\int_{-3\sigma}^{3\sigma}dE^{\prime}\ \sigma[(E^{\prime}+\frac{E_{n+1}+E_n}{2})^2]\left[\mathrm{Erf}\left(\frac{-0.01-E^{\prime}}{\sqrt{2}\sigma}\right)-\mathrm{Erf}\left(\frac{0.01-E^{\prime}}{\sqrt{2}\sigma}\right)\right].
    \label{eq:events}
\end{equation}
As the experimental data is the event distributions instead of the cross section distributions, we introduce the normalization factor $a_1=1$ and $a_2=a$ for the $\gamma\gamma\to D^0\bar{D}^0$ and $\gamma\gamma\to D^+D^-$ channels, respectively.
Finally, we use Eq.~\eqref{eq:events} to fit the experimental result~\cite{BaBar:2010jfn} and the best fit results are shown in Fig.\ref{fig:fit_FSI}, with $\chi^2/\mathrm{d.o.f}=2.04$.
We use the standard bootstrap method to derive the error band. In total, we generate 10,000 sample sets and perform a fit for each sample set to obtain 10,000 sets of parameters.
The distribution of the parameters is shown in Fig.\ref{fig:para_dis},
and the corresponding central values, as well as the errors, are presented in Tab.~\ref{tab:low-energy coe}.
\begin{table*}[htp!]
    \centering
    \caption{The parameters of the $\gamma\gamma\to D\bar{D}$ scattering.}
    \renewcommand{\arraystretch}{1.5}
    \resizebox{0.75\textwidth}{!}{
    \begin{tabular}{c|cccccc}
    \hline
       Parameters&$F_{A}\ [\mathrm{GeV^{-2}}]$  &$F_{A}^{\lambda}\ [\mathrm{GeV^{-2}}]$&$\Lambda\ [\mathrm{GeV}]$&$\mathcal{U}\ [\mathrm{GeV^{-2}}]$&$r\ [\mathrm{GeV^{0}}]$&$a\ [\mathrm{GeV^{-1}}]$\\
       \hline
       Value&$45.143\pm39.667$ & $-12.232\pm7.646$ & $1.942\pm0.251$& $307.547\pm33.112$& $0.823\pm0.150$& $0.284\pm 0.036$\\
        \hline
    \end{tabular}
    }
    \label{tab:low-energy coe}
\end{table*}
According the center values of parameters of Tab.~\ref{tab:low-energy coe}, the poles position of $T$-matrix are shown in Tab.~\ref{tab:pole_position}.
\begin{table*}[htp!]
    \centering
    \caption{The poles on various RSs are listed only for the RS that is closest to the physical observable real axis (where the real part corresponds to the energy and the imaginary part to half the width).}
    \renewcommand{\arraystretch}{1.5}
    \resizebox{0.75\textwidth}{!}{
    \begin{tabular}{c|@{\hspace{20pt}}c@{\hspace{20pt}}|@{\hspace{20pt}}c@{\hspace{20pt}}|@{\hspace{20pt}}c@{\hspace{20pt}}|@{\hspace{20pt}}c@{\hspace{20pt}}}
    \hline
       RSs&$+++$  &$-++$&$--+$&$---$\\
       \hline
       \multirow{2}{*}{$E$ [GeV]}&$3.65$&$3.76\pm0.003i$&$3.73$&$3.73$\\
       & & & $3.94\pm 0.003i$\\
        \hline
    \end{tabular}
    }
    \label{tab:pole_position}
\end{table*}

Using the parameters from Tab.~\ref{tab:low-energy coe}, the real parts of the T-matrix at $s=m_{B^0}^0$ and $s=m_{B_s^0}^0$ are given as 
\begin{equation}
   \mathrm{Re}[t(m_{B^0}^2)][\mathrm{GeV}^{-2}]= \begin{pmatrix}
0.9936\pm0.2974 & 0.1541\pm0.0326& 
  0.1593\pm0.0380\\
  0.1541\pm0.0326& 0.9943\pm0.2972 & 
0.1595\pm0.0383 \\ 0.1593\pm0.0380
& 0.1595\pm0.0383& 
  1.1346\pm0.2901
\end{pmatrix},
\label{eq:T_b_re}
\end{equation}
\begin{equation}
   \mathrm{Re}[t(m_{B_s}^2)][\mathrm{GeV}^{-2}]= \begin{pmatrix}
0.8938\pm0.2881 & 0.1425\pm0.0279& 
  0.1478\pm0.0317\\
  0.1425\pm0.0279& 0.8990\pm0.2880 & 
0.1480\pm0.0319 \\ 0.1478\pm0.0317
&  0.1480\pm0.0319& 
  1.0198\pm0.2846
\end{pmatrix}.
\label{eq:T_bs_re}
\end{equation}
The corresponding imaginary parts read as 
\begin{equation}
   \mathrm{Im}[t(m_{B^0}^2)][\mathrm{GeV}^{-2}]= \begin{pmatrix}
0.3753\pm0.2155 & 0.1260\pm0.0430& 
  0.1398\pm0.0443\\
  0.1260\pm0.0430& 0.3796\pm0.2169 & 
0.1405\pm0.0443 \\ 0.1398\pm0.0443
&  0.1405\pm0.0443& 
  0.4872\pm0.2489
\end{pmatrix},
\label{eq:T_b_im}
\end{equation}
\begin{equation}
   \mathrm{Im}[t(m_{B_s}^2)][\mathrm{GeV}^{-2}]= \begin{pmatrix}
0.3112\pm0.1915 & 0.1045\pm0.0393& 
  0.1158\pm0.0407\\
  0.1045\pm0.0393& 0.3147\pm0.1928 & 
0.1163\pm0.0407 \\ 0.1158\pm0.0407
&  0.1163\pm0.0407& 
  0.4023\pm0.2215
\end{pmatrix}.
\label{eq:T_bs_im}
\end{equation}
The numerical results of the two-body propagators  $G^{\Lambda}(m_{B^0}^2)$ and $G^{\Lambda}(m_{B_s}^2)$ read as
\begin{equation}
   \mathrm{Re}[G^{\Lambda}(m_{B^0}^2)][\mathrm{GeV}^{2}]= \begin{pmatrix}
-0.1436\pm0.0151 & 0& 
  0\\
  0& -0.1438\pm0.0150 & 
0 \\ 0
&  0& 
  -0.1489\pm0.0135
\end{pmatrix},
\label{eq:G_b_re}
\end{equation}
\begin{equation}
   \mathrm{Re}[G^{\Lambda}(m_{B_s}^2)][\mathrm{GeV}^{2}]= \begin{pmatrix}
-0.1445\pm0.0162 & 0& 
  0\\
  0& -0.1448\pm0.0162 & 
0 \\ 0
&  0& 
  -0.1506\pm0.0150
\end{pmatrix},
\label{eq:G_bs_re}
\end{equation}
\begin{equation}
   \mathrm{Im}[G^{\Lambda}(m_{B^0}^2)][\mathrm{GeV}^{2}]= \begin{pmatrix}
0.0446\pm0.0197 & 0& 
  0\\
  0& 0.0450\pm0.0198 & 
0 \\ 0&  0& 
  0.0539\pm0.0214
\end{pmatrix},
\label{eq:G_b_im}
\end{equation}
\begin{equation}
   \mathrm{Im}[G^{\Lambda}(m_{B_s}^2)][\mathrm{GeV}^{2}]= \begin{pmatrix}
0.0410\pm0.0192 & 0& 
  0\\
  0& 0.0414\pm0.0193 & 
0 \\ 0&  0& 
  0.0498\pm0.0211
\end{pmatrix}.
\label{eq:G_bs_im}
\end{equation}

\section{The effect of FSI for $\mathcal{A}_{\mathrm{CP}}$}
\label{app:FSI for acp}
From Tab.~\ref{tab:decay channel}, one can read the bare production amplitude for each channel. 
The bare production amplitudes for the 
$\bar{B}^0\to D^0\bar{D}^0$, 
$\bar{B}^0\to D^+D^-$, $\bar{B}^0\to D_s^+D_s^-$ processes can be rewritten as
\begin{equation}
    \begin{aligned}
        \mathcal{M}(\bar{B}^0\to D^0\bar{D}^0)&= Y_{cd} \mathcal{A} \exp{i\delta_a},\\
    \mathcal{M}(\bar{B}^0\to D^+D^-)&= Y_{cd} (\mathcal{T}+\mathcal{A}\exp(i\delta_a))-Y_{cd} R\exp(i\gamma) \mathcal{P}\exp(i\delta_p),\\
     \mathcal{M}(\bar{B}^0\to D_s^+D_s^-)&= Y_{cd} \mathcal{A} \exp{i\delta_a}.
    \end{aligned}
\end{equation}
Here $\mathcal{T}$, $\mathcal{A}$, and $\mathcal{P}$ are the absolute values of $T$, $A$ and $P$, respectively. $\delta_a$ ($\delta_p$) is the relative angle between the amplitudes $A$ ($P$) and $T$. $R\equiv |Y_{ud}/Y_{cd}|$ and $\gamma\equiv \arg(-Y_{ud}/Y_{cd})$. With these definitions, the amplitudes of their  corresponding CP related channels are 
\begin{equation}
    \begin{aligned}
        \mathcal{M}(B^0\to D^0\bar{D}^0)&= Y^*_{cd} \mathcal{A} \exp{i\delta_a},\\
    \mathcal{M}(B^0\to D^+D^-)&= Y_{cd}^* (\mathcal{T}+\mathcal{A}\exp(i\delta_a))-Y_{cd}^* R\exp(-i\gamma) \mathcal{P}\exp(i\delta_P),\\
     \mathcal{M}(B^0\to D_s^+D_s^-)&= Y_{cd}^* \mathcal{A} \exp{i\delta_a}.
    \end{aligned}
\end{equation}

The general form of the $T$-matrix reads as 
\begin{eqnarray}
    t=\begin{pmatrix}
        t_{11}\exp(i\theta_{11})& t_{12}\exp(i\theta_{12})& t_{13}\exp(i\theta_{13})\\
        t_{12}\exp(i\theta_{12})& t_{22}\exp(i\theta_{22})& t_{23}\exp(i\theta_{23})\\
        t_{13}\exp(i\theta_{13})& t_{23}\exp(i\theta_{23})& t_{33}\exp(i\theta_{33})\\
    \end{pmatrix}.
\end{eqnarray}
The two-body propagator is parameterized as 
\begin{eqnarray}
    G=\mathrm{diag}(g_1\exp(i\alpha_1),g_2\exp(i\alpha_2),g_3\exp(i\alpha_3)).
\end{eqnarray}
If we assume $\mathcal{P}\sim \mathcal{T}$ and $r\equiv \mathcal{A}/\mathcal{T}$, the $\mathcal{A}_{\mathrm{CP}}$s including FSI are given by,
\begin{equation}
    \begin{aligned}
        \mathcal{A}_{CP}(\bar{B}^0\to D^0\bar{D}^0)&=\frac{2R\sin{\gamma}}{1+R^2-2R\cos{\gamma}}\left[\delta_p+\frac{g_1t_{11}}{g_2t_{12}}r\sin{(\alpha_1-\alpha_2+\theta_{11}-\theta_{12})}-\frac{g_3t_{13}}{g_2t_{12}}r\sin{(\alpha_2-\alpha_3+\theta_{12}-\theta_{13})}\right. \\
        &\left.-\frac{r\sin{(\alpha_2+\theta_{12})}}{g_2t_{12}}\right]+\mathcal{O}(r^2)+\mathcal{O}(\delta_p^2)+\mathcal{O}(r,\delta_p)+\mathcal{O}(r,\delta_a-\pi),\\
        \mathcal{A}_{CP}(\bar{B}^0\to D^+D^-)&=\frac{2R\sin{\gamma}}{1+R^2-2R\cos{\gamma}}\left\{\delta_p+\frac{g_1t_{12}[\sin{(\alpha_1+\theta_{12})}+g_2t_{22}\sin{(\alpha_1-\alpha_2+\theta_{12}-\theta_{22})}]r}{1+g_2t_{22}(2\cos{(\alpha_2+\theta_{22})+g_2t_{22}})}\right. \\
        +&\left.\frac{g_3t_{23}[\sin{(\alpha_3+\theta_{23})}-g_2t_{22}\sin{(\alpha_2-\alpha_3+\theta_{22}-\theta_{23})}]r}{1+g_2t_{22}(2\cos{(\alpha_2+\theta_{22})+g_2t_{22}})}\right\}+\mathcal{O}(r^2)+\mathcal{O}(\delta_p^2)+\mathcal{O}(r,\delta_p)+\mathcal{O}(r,\delta_a-\pi),\\
        \mathcal{A}_{CP}(\bar{B}^0\to D_s^+D_s^-)&=\frac{2R\sin{\gamma}}{1+R^2-2R\cos{\gamma}}\left[\delta_p+\frac{g_1t_{13}}{g_2t_{23}}r\sin{(\alpha_1-\alpha_2+\theta_{13}-\theta_{23})}-\frac{g_3t_{33}}{g_2t_{23}}r\sin{(\alpha_2-\alpha_3+\theta_{23}-\theta_{33})}\right. \\
        &\left.-\frac{r\sin{(\alpha_2+\theta_{23})}}{g_2t_{23}}\right]+\mathcal{O}(r^2)+\mathcal{O}(\delta_p^2)+\mathcal{O}(r,\delta_p)+\mathcal{O}(r,\delta_a-\pi).
    \end{aligned}
\end{equation}
If we ignore the FSI, the $\mathcal{A}_{\mathrm{CP}}$ for the three channels denoted as
\begin{equation}
    \begin{aligned}
        \mathcal{A}^{\prime}_{CP}(\bar{B}^0\to D^0\bar{D}^0)&=0,\\
        \mathcal{A}^{\prime}_{CP}(\bar{B}^0\to D^+D^-)&=\frac{2R\delta_p\sin{\gamma}}{1+R^2-2R\cos{\gamma}}+\mathcal{O}(r^2)+\mathcal{O}(\delta_p^2)+\mathcal{O}(r,\delta_p)+\mathcal{O}(r,\delta_a-\pi),\\
        \mathcal{A}_{CP}^{\prime}(\bar{B}^0\to D_s^+D_s^-)&=0.
    \end{aligned}
\end{equation}
With the definitions
\begin{equation}
    \begin{aligned}
        &f\equiv \frac{2R\sin{\gamma}}{1+R^2-2R\cos{\gamma}},\\
        &\Delta_{jk}^{i}\equiv \frac{g_jt_{ij}}{g_kt_{ik}}r\sin{(\alpha_j-\alpha_k+\theta_{ij}-\theta_{ik})},\quad\quad\quad j\ne k,\\
        &\Delta_{jk}^{i}\equiv \frac{1}{g_jt_{ij}}r\sin{(\alpha_j+\theta_{ij})},\quad\quad\quad\quad\quad j=k,\\
        &\Delta^{\prime}_{ij}\equiv \frac{g_{i}t_{ij}}{1+g_{j}t_{jj}[g_jt_{jj}+2\cos{(\alpha_j+\theta_{jj})}]}r[\sin{(\alpha_i+\theta_{ij})}+g_jt_{jj}\sin{(\alpha_i-\alpha_j+\theta_{ij}-\theta_{jj})}],
    \end{aligned}
\end{equation}
the corrections caused by FSI to  $\mathcal{A}_{\mathrm{CP}}$ are 
\begin{equation}
    \begin{aligned}
        &\Delta_1=f(\delta_p+\Delta_{12}^1+\Delta_{32}^{1}-\Delta_{22}^{1})+\mathcal{O}(\cdots),\\
        &\Delta_2=f(\Delta_{12}^{\prime}+\Delta_{32}^{\prime})+\mathcal{O}(\cdots),\\
        &\Delta_3=f(\delta_p+\Delta_{12}^3+\Delta_{32}^{3}-\Delta_{22}^{3})+\mathcal{O}(\cdots),
    \end{aligned}    
\end{equation}
for each channel. 

\section{Bootstrap}
\label{app:bootstrap}
\begin{figure}[htb!]
    \centering
    \includegraphics[width=0.76\linewidth]{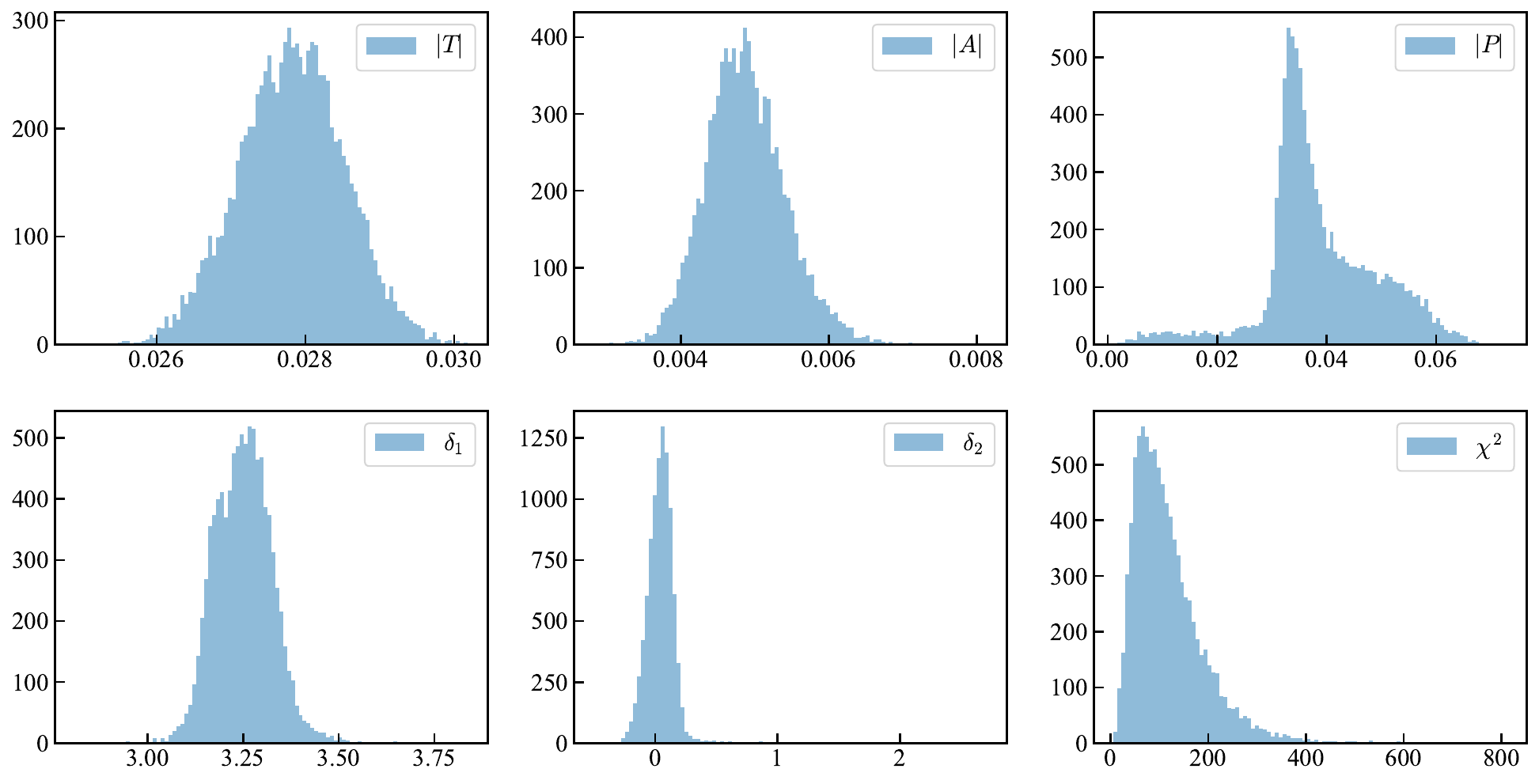}
    \includegraphics[width=0.76\linewidth]{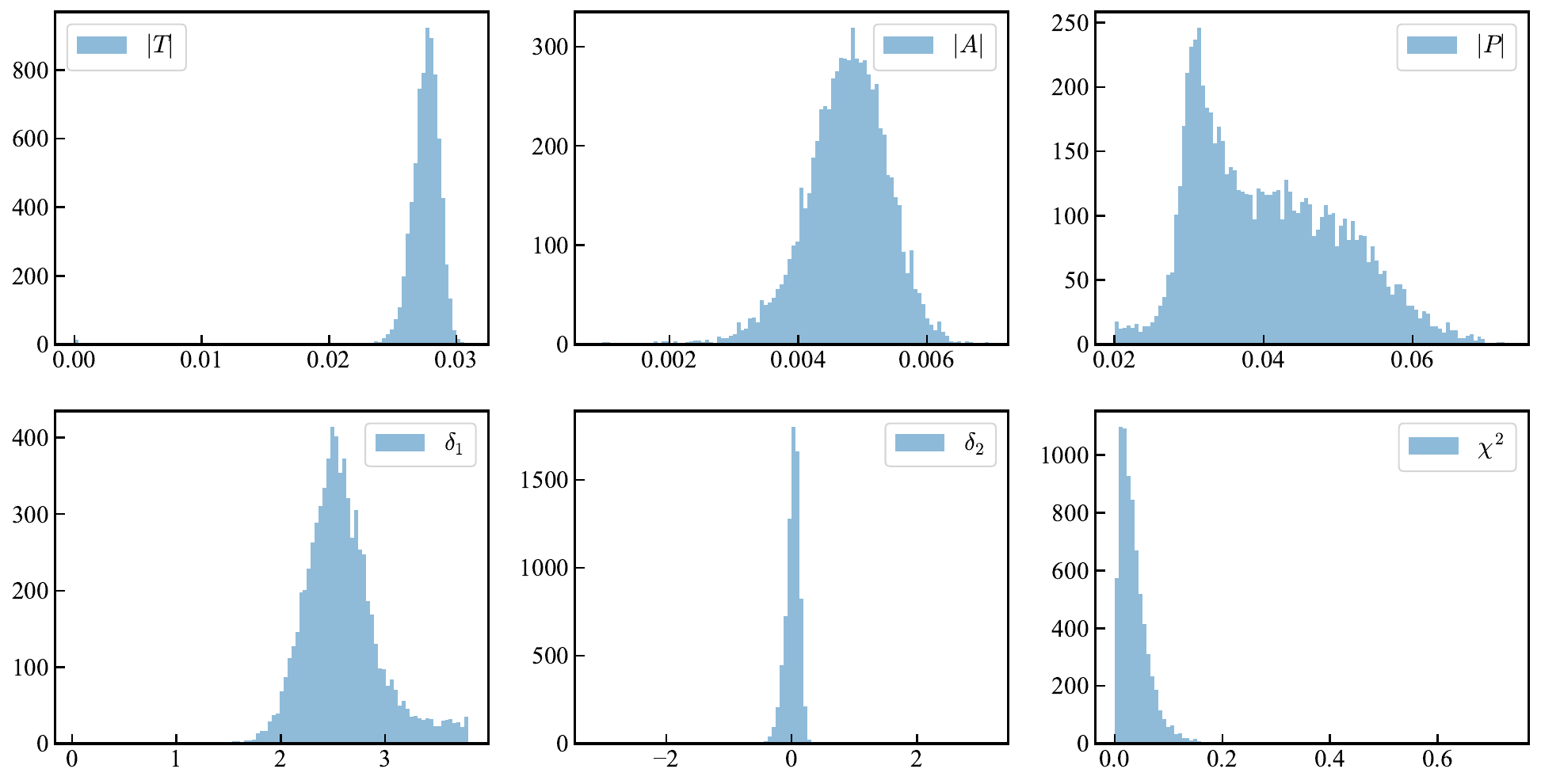}
    \includegraphics[width=0.76\linewidth]{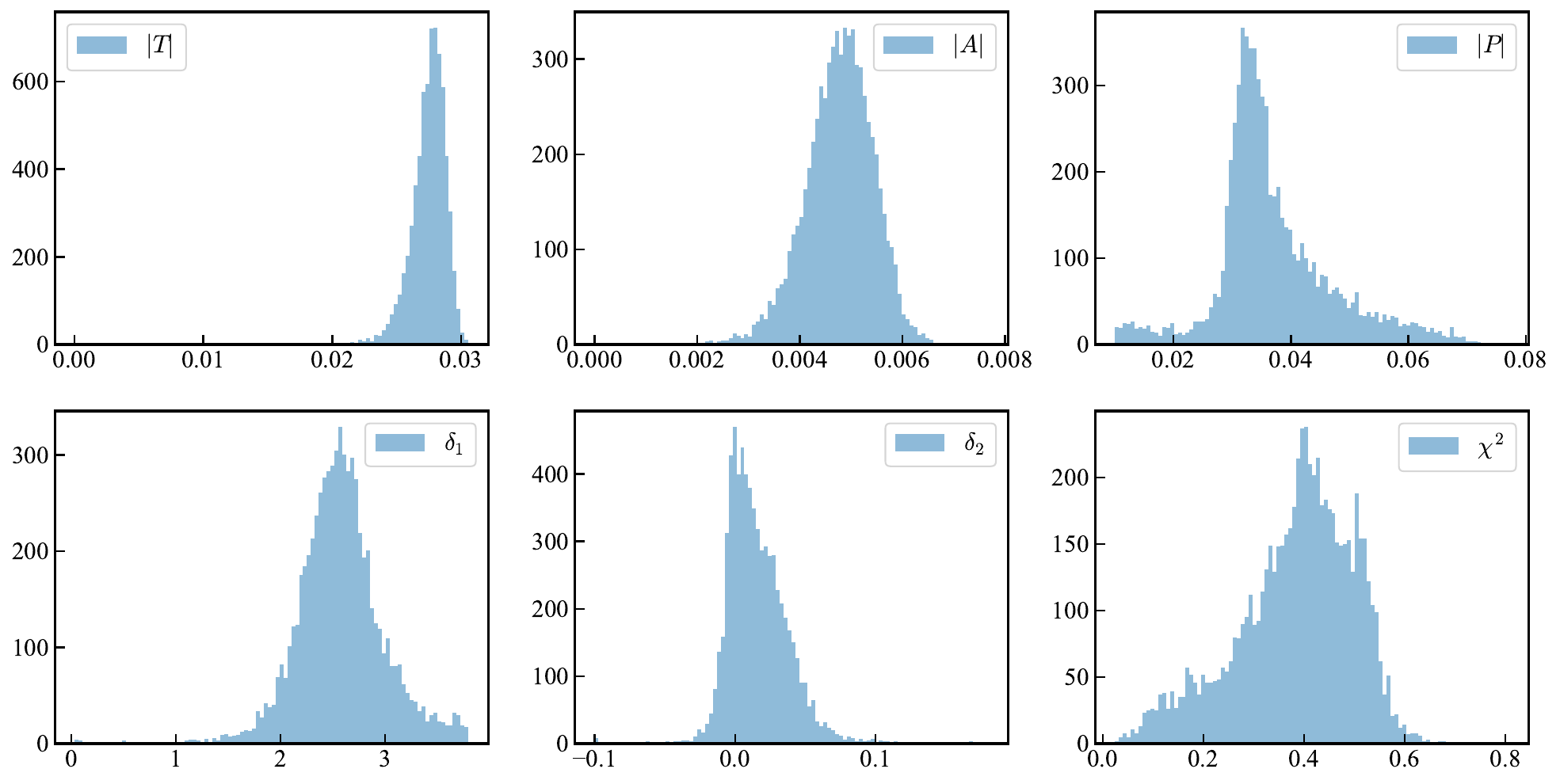}
    \caption{Distributions of weak decays parameters. The upper, middle and lower panels correspond to scheme I, scheme II, and scheme III in Tab.~\ref{tab:fit_para}.}
    \label{fig:para_weak_dis}
\end{figure}

The errors of the fitted parameters are derived by making bootstrap from three fit schemes, respectively. From the experimental data within the experimental error range, 10,000 sample sets are generated for each fit scheme and three fit schemes are performed to derive 10,000 sets of parameters for the corresponding scheme. Since not all of the 10,000 fits converge well, one needs to select reasonable parameter range. Finally, we select 9,445 sets of parameters to derive the parameter errors for scheme I, 7381 sets for scheme II, and 6788 sets for scheme III. The distribution of the parameters of the three fit schemes are shown in Fig.~\ref{fig:para_weak_dis}.


\section{Numerical results and experimental values}
\label{app:results and exp}
In Tab.~\ref{tab:width_Bbar} and Tab.~\ref{tab:width_B}, we listed the width that  $\bar{B}$ and $B$ decay to two charmed mesons for theoretical results of scheme II and scheme III and experimental values. 
The absolute values of the amplitudes $T$, $P$ and $A$ in the three schemes are almost the same and consistent with the pQCD and pole-model estimation.
In pQCD factorization with factorization scale $\mu_f=m_b$, the emission diagram can be parameterized as\cite{Zhou:2015jba,Zhou:2024qmm}
\begin{equation}
    \begin{aligned}
        E&\sim \frac{G_F}{\sqrt{2}}a_{2}F^{B\to D}(m_D^2)f_D(m_B^2-m_D^2),
    \end{aligned}
\end{equation}
where
\begin{equation}
    F^{B\to D}(m_D^2)=\frac{F^{B\to D}(0)}{1-\alpha_1\frac{m_D^2}{m_{\mathrm{pole}}^2}+\alpha_2\frac{m_D^4}{m_{\mathrm{pole}}^4}},
\end{equation}
with $\alpha_1=1.71$, $\alpha_2=0.52$~\cite{Zhou:2015jba}, and $m_{\mathrm{pole}}=m_B$.
The annihilation diagram cannot be precisely calculated from quark-level, but can be estimated by
Pole-dominance model~\cite{Fu-Sheng:2011fji}
\begin{equation}
    \begin{aligned}
        A&\sim \frac{G_F}{\sqrt{2}}a_{1}f_Bf_{\chi_{c0}}g_{\chi_{c0} D\bar{D}} \frac{m_B^2}{m_B^2-m_{\chi_{c0}}^2},
    \end{aligned}
\end{equation}
where $a_{1,2}=C_{1,2}+\frac{C_{2,1}}{N_c}=0.090,1.036$ are combinations of Wilson coefficients and $f_D=(212\pm 7)~\mathrm{MeV}$, $f_B=(190\pm 1.3)~\mathrm{MeV}$ are $D$ and $B$ meson decay constant.
The $\chi_{c0}$ decay constant  $f_{\chi_{c0}}=(510\pm 40)~\mathrm{MeV}$, and its coupling $g_{\chi_{c0}D\bar{D}}=(-25.036\pm 1.964)\mathrm{GeV}$ to $D\bar{D}$ can be found in Ref.~\cite{Colangelo:2003sa,Colangelo:2002mj}. The $\langle D|(\bar{q}b)_{V-A}|B\rangle$ transition form factor
$F^{B\to D}(0)=0.54$~\cite{Ou-Yang:2025ije} at zero momentum transfer is used.
Consequently, we can estimate the absolute values of $E$ and $A$ about $0.03~\mathrm{MeV}$ and $0.003~\mathrm{MeV}$. These values are at the same order of magnitude as our results.
\begin{table*}[htb!]
  \centering
    \renewcommand{\arraystretch}{1.5}
    \caption{The width of $\bar{B}$ for fit scheme II, scheme III, and experiment. The lifetime of $\bar{B}^0$, $\bar{B}_s^0$, and $B^-$ are given by Ref.~\cite{Belle-II:2023bps,LHCb:2017knt, LHCb:2014qsd} where we ignore the lifetime difference of $B$ and $\bar{B}$.}
    \resizebox{0.96\textwidth}{!}{
    \begin{tabular}{c|@{\hspace{20pt}}c@{\hspace{40pt}}c@{\hspace{20pt}}@{\hspace{20pt}}c@{\hspace{40pt}}c@{\hspace{40pt}}c}
        \hline
        Channel & II & II w.o. FSI & III & III w.o. FSI & Experiment\\
       \hline
        $\bar{B}^0\to D^0\bar{D}^0$& $(4.711\pm1.261)\times 10^{-15}$& $(5.392\pm1.501)\times 10^{-15}$&$(4.700\pm1.273)\times 10^{-15}$ & $(5.482\pm1.542)\times 10^{-15}$ & $(6.147\pm 2.913)\times 10^{-15}$~\cite{LHCb:2013sad}\\
        \hline
        $\bar{B}^0\to D^+D^-$& $(7.651\pm2.562)\times 10^{-14}$ & $(1.053\pm 0.339)\times 10^{-13}$ &$(7.355\pm 1.374)\times 10^{-14}$& $(1.010\pm 0.157)\times 10^{-13}$ & $\cdots$\\
        \hline
        $\bar{B}^0\to D_s^+D_s^-$&$(4.187\pm 1.136)\times 10^{-15}$  & $(5.076\pm 1.413)\times 10^{-15}$&$(4.166\pm1.145)\times 10^{-15}$ & $(5.161\pm1.452)\times 10^{-15}$ & $<(1.581\pm 0.161)\times 10^{-14}$~\cite{Zupanc:2007pu} \\
        \hline
        $\bar{B}_s^0\to D^0\bar{D}^0$& $(8.498\pm 2.456)\times 10^{-14}$ & $(1.010\pm 0.281)\times 10^{-13}$&$(8.523\pm2.507)\times 10^{-14}$& $(1.027\pm0.289)\times 10^{-13}$ & $(8.084\pm 2.213)\times 10^{-14}$~\cite{LHCb:2013sad}\\
        \hline
        $\bar{B}_s^0\to D^+D^-$& $(8.456\pm 2.447)\times10^{-14}$ & $(1.008\pm 0.280)\times10^{-13}$ &$(8.480\pm 2.497)\times 10^{-14}$ & $(1.024\pm 0.288)\times 10^{-13}$ & $(9.360\pm 2.294)\times 10^{-14}$ ~\cite{LHCb:2013sad}\\
        \hline
        $\bar{B}_s^0\to D_s^+D_s^-$& $(1.723\pm 0.343)\times 10^{-12}$ & $(2.459\pm 0.412)\times 10^{-12}$&$(1.716\pm 0.326)\times 10^{-12}$& $(2.449\pm 0.402)\times 10^{-12}$ & $(1.702\pm0.230)\times 10^{-12}$~\cite{LHCb:2013sad}\\
        \hline
        $B^-\to D^0D^-$& $(1.585\pm 0.409)\times 10^{-13}$ & $(1.585\pm 0.409)\times 10^{-13}$ &$(1.536\pm 0.178)\times 10^{-13}$ & $(1.536\pm 0.178)\times 10^{-13}$ & $\cdots$\\
        \hline
        $B^-\to D^0D_s^-$&$(3.396\pm 0.395)\times 10^{-12}$ &$(3.396\pm 0.395)\times 10^{-12}$ &$(3.363\pm 0.364)\times10^{-12}$& $(3.363\pm 0.364)\times10^{-12}$ & $(3.820\pm 0.316)\times 10^{-12}$~\cite{Belle-II:2024xtf}\\
        \hline
        $\bar{B}^0\to D^+D_s^-$& $(3.391\pm0.395)\times 10^{-12}$ & $(3.391\pm0.395)\times 10^{-12}$&$(3.358\pm 0.363)\times 10^{-12}$& $(3.358\pm 0.363)\times 10^{-12}$ & $(3.908\pm 0.313)\times 10^{-12}$~\cite{Belle-II:2024xtf}\\
        \hline
        $\bar{B}^0_s\to D_s^+D^-$& $(1.541\pm 0.398)\times 10^{-13}$ &  $(1.541\pm 0.398)\times 10^{-13}$&$(1.494\pm 0.173)\times 10^{-13}$ &$(1.494\pm 0.173)\times 10^{-13}$ & $(1.532\pm 0.333)\times 10^{-13}$~\cite{LHCb:2013sad}\\
        \hline
    \end{tabular}
    }
    \label{tab:width_Bbar}
\end{table*}

\begin{table*}[htb!]
    \centering
    \renewcommand{\arraystretch}{1.5}
    \caption{The width of $B$ for fit scheme II, scheme III, and experiment. The lifetime of $B^0$, $B_s^0$, and $B+$ are given by Ref.~\cite{Belle-II:2023bps,LHCb:2017knt, LHCb:2014qsd} where we ignore the lifetime difference of $B$ and $\bar{B}$.}
    \resizebox{0.96\textwidth}{!}{
    \begin{tabular}{c|@{\hspace{20pt}}c@{\hspace{40pt}}c@{\hspace{20pt}}@{\hspace{20pt}}c@{\hspace{40pt}}c@{\hspace{40pt}}c}
        \hline
        Channel & II & II w.o. FSI & III & III w.o. FSI & Experiment\\
       \hline
        $B^0\to D^0\bar{D}^0$& $(3.435\pm1.176)\times 10^{-15}$& $(5.392\pm1.501)\times 10^{-15}$&$(3.508\pm1.204)\times 10^{-15}$ & $(5.482\pm1.542)\times 10^{-15}$ & $\cdots$\\
        \hline
        $B^0\to D^+D^-$& $(9.455\pm3.116)\times 10^{-14}$ & $(1.318\pm 0.418)\times 10^{-13}$ &$(9.273\pm 2.087)\times 10^{-14}$& $(1.292\pm 0.266)\times 10^{-13}$ & $(9.309\pm 1.062)\times10^{-14}$~\cite{Belle:2012mef}\\
        \hline
        $\bar{B}^0\to D_s^+D_s^-$&$(2.924\pm 1.049)\times 10^{-15}$  & $(5.076\pm 1.413)\times 10^{-15}$&$(2.989\pm1.079)\times 10^{-15}$ & $(5.161\pm1.452)\times 10^{-15}$ & $\cdots$ \\
        \hline
        $B_s^0\to D^0\bar{D}^0$& $(8.624\pm 2.461)\times 10^{-14}$ & $(1.010\pm 0.281)\times 10^{-13}$&$(8.641\pm2.512)\times 10^{-14}$& $(1.027\pm0.289)\times 10^{-13}$ & $\cdots$\\
        \hline
        $B_s^0\to D^+D^-$& $(8.583\pm 2.452)\times10^{-14}$ & $(1.008\pm 0.280)\times10^{-13}$ &$(8.599\pm 2.502)\times 10^{-14}$ & $(1.024\pm 0.288)\times 10^{-13}$ & $\cdots$ \\
        \hline
        $B_s^0\to D_s^+D_s^-$& $(1.707\pm 0.337)\times 10^{-12}$ &$(2.434\pm 0.403)\times 10^{-12}$ &$(1.698\pm 0.330)\times 10^{-12}$& $(2.422\pm 0.393)\times 10^{-12}$ & $\cdots$\\
        \hline
        $B^+\to \bar{D}^0D^+$& $(1.501\pm 0.397)\times 10^{-13}$ & $(1.501\pm 0.397)\times 10^{-13}$ &$(1.482\pm 0.171)\times 10^{-13}$ & $(1.536\pm 0.178)\times 10^{-13}$ & $(1.548\pm 0.198)\times 10^{-13}$~\cite{Belle:2008doh}\\
        \hline
        $B^+\to \bar{D}^0D_s^+$&$(3.404\pm 0.396)\times 10^{-12}$ & $(3.404\pm 0.396)\times 10^{-12}$ &$(3.368\pm 0.364)\times10^{-12}$& $(3.368\pm 0.364)\times10^{-12}$ & $\cdots$\\
        \hline
        $B^0\to D^-D_s^+$& $(3.399\pm0.395)\times 10^{-12}$ & $(3.399\pm0.395)\times 10^{-12}$ &$(3.363\pm 0.363)\times 10^{-12}$& $(3.363\pm 0.363)\times 10^{-12}$ & $\cdots$\\
        \hline
        $B^0_s\to D_s^-D^+$& $(1.460\pm 0.386)\times 10^{-13}$ & $(1.460\pm 0.386)\times 10^{-13}$ &$(1.441\pm 0.166)\times 10^{-13}$ &$(1.441\pm 0.166)\times 10^{-13}$ & $\cdots$\\
        \hline
    \end{tabular}
    }
    \label{tab:width_B}
\end{table*}

In Tab.~\ref{tab:acp_II&III}, we list the $\mathcal{A}_{\mathrm{CP}}$ for the $\bar{B}$ meson decay to a pair of charmed mesons in  scheme II and scheme III, with a comparison with experimental values.
\begin{table*}[htb!]
    \centering
    \renewcommand{\arraystretch}{1.5}
    \caption{The $\mathcal{A}_{\mathrm{CP}}$ for fit scheme II, scheme III, and experiment.}
    \resizebox{0.96\textwidth}{!}{
    \begin{tabular}{c|@{\hspace{20pt}}c@{\hspace{40pt}}c@{\hspace{20pt}}@{\hspace{20pt}}c@{\hspace{40pt}}c@{\hspace{40pt}}c}
        \hline
        Channel & II & II w.o. FSI & III & III w.o. FSI & Experiment\\
       \hline
        $\bar{B}^0\to D^0\bar{D}^0$&  $(1.566\pm 0.849)\times10^{-1}$ & $0\pm 0$&$(1.452\pm 0.762)\times 10^{-1}$& $0\pm0$ & $\cdots$\\
        \hline
        $\bar{B}^0\to D^+D^-$& $(-1.055\pm2.643)\times 10^{-1}$& $(-1.118\pm2.688)\times10^{-1}$ &$(-1.154\pm 0.756)\times10^{-1}$& $(-1.225\pm 0.829)\times10^{-1}$ & $-0.128\pm 0.103~(\mathrm{stat})\pm 0.010~(\mathrm{syst})$~\cite{LHCb:2024gkk}\\
        \hline
        $\bar{B}^0\to D_s^+D_s^-$ & $(1.776\pm 0.931)\times 10^{-1}$& $0\pm 0$ &$(1.646\pm 0.841)\times 10^{-1}$& $0\pm0$ & $\cdots$\\
        \hline
        $\bar{B}_s^0\to D^0\bar{D}^0$&  $(-7.382\pm 4.243)\times10^{-3}$ & $0\pm 0$&$(-6.892\pm3.775)\times 10^{-3}$ & $0\pm 0$ & $\cdots$\\
        \hline
        $\bar{B}_s^0\to D^+D^-$& $(-7.435\pm 4.267)\times 10^{-3}$ & $0\pm 0$ &$(-6.944\pm 3.798)\times 10^{-3}$ & $0\pm 0$ & $\cdots$\\
        \hline
        $\bar{B}_s^0\to D_s^+D_s^-$& $(0.485\pm1.216)\times 10^{-2}$ &  $(0.513\pm 1.241)\times 10^{-2}$&$(5.184\pm 3.502)\times 10^{-3}$ & $(5.482\pm 3.815)\times 10^{-3}$ & $0.135\pm0.108~(\mathrm{stat})\pm 0.027~(\mathrm{syst})$~\cite{LHCb:2024gkk}\\
        \hline
        $B^-\to D^0D^-$& $(0.272\pm 2.311)\times 10^{-1}$ &$(0.272\pm 2.311)\times 10^{-1}$ &$(1.787\pm 2.574)\times 10^{-2}$ & $(1.787\pm 2.574)\times 10^{-2}$& $(2.3\pm 2.7\pm 0.4)\times 10^{-2}$~\cite{LHCb:2018uli}\\
        \hline
        $B^-\to D^0D_s^-$& $(-0.120\pm 1.020)\times 10^{-2}$ &  $(-0.120\pm 1.020)\times 10^{-2}$&$(-0.779\pm1.126)\times 10^{-3}$ & $(-0.779\pm1.126)\times 10^{-3}$ & $(-0.4\pm0.5\pm0.5)\times 10^{-2}$~\cite{LHCb:2018uli}\\
        \hline
        $\bar{B}^0\to D^+D_s^-$& $(-0.120\pm 1.020)\times 10^{-2}$ & $(-0.120\pm 1.020)\times 10^{-2}$&$(-0.779\pm1.126)\times 10^{-3}$ &$(-0.779\pm1.126)\times 10^{-3}$ & $(0.9\pm 5.3\pm 4.0)\times10^{-3}$~\cite{LHCb:2026lvd}\\
        \hline
        $\bar{B}^0_s\to D_s^+D^-$& $(0.272\pm 2.311)\times 10^{-1}$ & $(0.272\pm 2.311)\times 10^{-1}$&$(1.787\pm 2.574)\times 10^{-2}$ & $(1.787\pm 2.574)\times 10^{-2}$ &  $0.103\pm 0.053\pm 0.010$~\cite{LHCb:2026lvd}\\
        \hline
    \end{tabular}
    }
    \label{tab:acp_II&III}
\end{table*}

\begin{table*}[htb!]
    \centering
    \caption{Fitted parameters in the three schemes. Errors are obtained with bootstrap resampling.}
    \resizebox{0.96\textwidth}{!}{
    \begin{tabular}{c|@{\hspace{15pt}}c@{\hspace{20pt}}c@{\hspace{20pt}}c@{\hspace{20pt}}c@{\hspace{20pt}}c@{\hspace{20pt}}c}
       \hline
       Scheme & $|T|~[\mathrm{MeV}]$& $|A|~[\mathrm{MeV}]$ & $|P|~[\mathrm{MeV}]$& $\delta_a[\mathrm{Rad}]$ & $\delta_p[\mathrm{Rad}]$& $\chi^2/\mathrm{d.o.f}$\\
        \hline
        I&$(2.78\pm 0.07)\times10^{-2}$& $(4.88\pm 0.54)\times10^{-3}$&$(3.86\pm 1.05)\times10^{-2}$& $3.25\pm0.08$& $(0.52\pm 1.54)\times 10^{-1}$& 2.30\\
        II&$(2.76\pm 0.16)\times10^{-2}$& $(4.74\pm 0.66)\times10^{-3}$&$(4.06\pm 0.99)\times 10^{-2}$& $2.59\pm 0.37$& $(0.23\pm 1.91)\times10^{-1}$& 1.15\\
        III&$(2.74\pm 0.14)\times10^{-2}$& $(4.78\pm 0.67)\times10^{-3}$&$(3.69\pm 0.99)\times10^{-2}$& $2.55\pm0.38$& $(1.60\pm 2.27)\times 10^{-2}$& 1.00\\
        \hline
    \end{tabular}
    }
    \label{tab:fit_para}
\end{table*}

\end{document}